\documentclass{emulateapj}

\newcommand{\HII}{H{\footnotesize II}} 
\newcommand{\HI}{H{\footnotesize I}} 

\shorttitle{Emerging Star Clusters in NGC 4449}
\shortauthors{Reines, Johnson \& Goss}

\begin{document}
 
\title{Emerging Massive Star Clusters Revealed:
High Resolution Imaging of NGC 4449 from the Radio to the Ultraviolet}

\author{Amy E. Reines and Kelsey E. Johnson}
\affil{Department of Astronomy, University of Virginia,
Charlottesville, VA, 22904-4325}
\email{areines@virginia.edu}

\and

\author{W. M. Goss}
\affil{National Radio Astronomy Observatory, P.O. Box O, 1003 Lopezville Rd,
Socorro, NM 87801}

\begin{abstract}
 
We present a multi-wavelength study of embedded massive clusters
in the nearby (3.9 Mpc) starburst galaxy NGC~4449 in an effort to
uncover the earliest phases of massive cluster evolution. By
combining high resolution imaging from the radio to the
ultraviolet, we reveal these clusters in the process of
emerging from their gaseous and dusty birth cocoons.  We
use Very Large Array (VLA) observations at centimeter wavelengths
to identify young clusters surrounded by ultra-dense \HII\
regions, detectable via their production of thermal free-free
radio continuum.  Ultraviolet, optical and infrared observations
are obtained from the {\it Hubble} and {\it Spitzer Space Telescope}
archives for comparison.  We detect 39 compact radio sources towards
NGC~4449 at 3.6 cm using the highest resolution (1\farcs3) and
sensitivity (RMS $\sim$ 12~$\mu$Jy) VLA image of the galaxy to date.
We reliably identify 13 thermal radio sources and their physical
properties are derived using both nebular emission from the \HII\
regions and SED fitting to the stellar continuum.  These radio detected
clusters have ages $\lesssim$~5~Myr and stellar masses of order
$10^4$~M$_\odot$.  The {\it measured} extinctions are quite low: 12 of the 13
thermal radio sources have $A_{\rm V} \lesssim 1.5$, while the most
obscured source has $A_{\rm V} \approx 4.3$.  By combining results from
the nebular and stellar emission, we find an {\it I}-band excess that
is anti-correlated with cluster age and an apparent mass-age correlation.
Additionally, we find evidence that local processes such as supernovae
and stellar winds are likely playing an important role in triggering
the current bursts of star formation within NGC~4449.
 
\end{abstract}


\section{Introduction}

\begin{deluxetable*}{ccccc}[t!] 
\tabletypesize{\footnotesize}
\tablecolumns{5} 
\tablewidth{0pt} 
\tablecaption{VLA Observations of NGC~4449\label{vlaobs}} 
\tablehead{ 
\colhead{Wavelength} & \colhead{Weighting}  & \colhead{Orig. Synthesized Beam\tablenotemark{a}}
&  \colhead{P.A.} & \colhead{RMS Noise} \\
\colhead{(cm)} & \colhead{(robust value)} & \colhead{(arcsec $\times$ arcsec)}
& \colhead{(degrees)} & \colhead{($\mu$Jy beam$^{-1}$)}}
\startdata
6.0 & 5 & 0.70 $\times$ 0.65 &  83 & 36 \\
3.6 & 1 & 1.27 $\times$ 1.14 & -73 & 12 \\
1.3 & 5 & 1.29 $\times$ 1.04 & -86 & 25
\enddata 
\tablenotetext{a}{The data have been
convolved to identical beams of 1\farcs3 $\times$
1\farcs3 to better facilitate the comparison between wavelengths.}
\end{deluxetable*} 

Young massive star clusters represent a fundamental mode of star formation
throughout the universe, both locally and at high redshift.  The most massive
and dense clusters are consistent with being young analogues of ancient
globular clusters like those in our own Milky Way \citep[e.g.][]{Ho96a,Ho96b}.
The prevalence of globular clusters around massive galaxies today suggests that these extreme
star clusters were forming ubiquitously in the early universe as
galaxies began to merge and coalesce hierarchically.  

Young massive clusters (YMCs) continue to play an important role
in the evolution of the universe.  YMCs and their constituent massive
stars ionize the interstellar medium, power the infrared radiation
of dust, produce heavy elements, and trigger further star formation via
supernova explosions.  In addition, the most massive YMCs, known as
``super star clusters'' \citep[see][and reference therein for overviews]{Whitmore03,OConnell04},
are packed with hundreds to
thousands of O and B stars.  These extreme clusters can drastically alter the
morphology of their host galaxy when the massive stars collectively explode at the
end of their lives, expelling huge amounts of gas and blowing galactic-scale
superbubbles \citep[e.g.][]{Tenorio07,Marlowe95}.

Despite the importance of massive stars and massive clusters, their formation
and earliest evolutionary stages are not well understood \citep{Tan05,Zinnecker07}.
We do know, however, that nearly all massive stars form in clusters and that YMCs host
large numbers of massive stars \citep{Hunter1999,Clark05,deWit05}.  Therefore, in addition
to being interesting in their own right, YMCs provide important laboratories for studying
the clustered mode of massive star formation, a crucial part of understanding massive star
formation in general. 

Interest in massive extragalactic star clusters was sparked by the discovery of a
population of YMCs in the peculiar galaxy NGC~1275 by \citet{Holtzmann91} using the
{\it Hubble Space Telescope (HST)}.  Since then, young massive clusters have
been found in a large number of galaxy systems, primarily at optical wavelengths
with {\it HST} \citep[for reviews, see][]{Whitmore03,Larsen06}.  More recently,
ultra-young ($\lesssim$~few Myr) massive clusters, still deeply embedded in their
birth material, have been discovered using infrared and radio imaging
\citep[e.g.][]{Kobulnicky99,Turner00,Neff00,Johnson01,Beck02,Johnson03,Johnson03b,
Johnson04,Tsai06}.  By combining observations spanning multiple wavelength regimes,
we aim to improve the current understanding of massive cluster (and massive star)
formation and early evolution.

Here we present our study of embedded massive clusters in the nearby starburst galaxy
NGC~4449.  NGC~4449 is a barred Magellanic-type irregular galaxy \citep{deVauc91}
with star formation occurring throughout the galaxy at a rate almost
twice that of the Large Magellanic Cloud \citep{Thronson87,Hunter99}.
NGC~4449 has almost certainly undergone an external perturbation, as is evident
in \HI\ studies revealing extended streamers wrapping around the galaxy and
counter-rotating (inner and outer) gas systems \citep[][and references therein]{Hunter98,Theis01}.
The proximity of NGC~4449 is advantageous for extragalactic star cluster research since
relatively small spatial scales can be probed.  The work of \citet{Gelatt01} has already
revealed $\sim 60$ optically-selected compact star clusters.  We adopt a distance of 3.9 Mpc,
consistent with recent work by \citet{Annibali07},
and thus 1\arcsec=19~pc.  

Using high-resolution radio imaging, we identify the youngest
embedded star clusters in NGC~4449.  The observed ultra-dense
\HII~regions are produced when hot massive stars within the clusters ionize the
surrounding gas and are identified as compact free-free (thermal) radio sources.
A previous catalog of radio detected \HII\ regions in
NGC~4449 was presented by \citet{Israel} using an NRAO 3-element
image at 11~cm (2.7~GHz) from \citet{Seaquist78} with a beam size of
$10\farcs8 \times 6\farcs5$ and an RMS noise of 0.4--0.6 mJy~beam$^{-1}$.
\citet{Israel} derived a list of eight \HII\ regions that had counterparts in
the optical H$\alpha$ data of \citet{Crillon69}.  The optical sizes were
10\arcsec--20\arcsec~and the 11~cm flux densities were 1--10 mJy.  
\citet{Israel80} also calculated the radio to H$\alpha$ derived extinction for seven
objects and compared these values with the Balmer decrement derived extinctions for
a single object (CM 39). 

Compared to the NRAO 3-element image at 11~cm used by \citet{Israel}, the current Very Large
Array (VLA) survey at 3.6 cm has a sensitivity that is improved by a factor of
$\sim$~40 and a beam solid angle that is smaller by a factor of $\sim$~40.
In the present work, we combine high-quality VLA data and archival images from the {\it Hubble}
and {\it Spitzer Space Telescopes} with stellar population synthesis
models in order to discern the physical properties of the radio-detected star clusters in NGC~4449
and gain insight into the earliest evolutionary phases experienced by massive star clusters.

\section{Observations}

\subsection{Radio Imaging with the VLA}\label{radioimaging}

Radio observations of NGC~4449 were obtained with the Very Large
Array\footnote{The National Radio Astronomy Observatory is a facility of
the National Science Foundation operated under cooperative agreement by
Associated Universities, Inc.} from April 2001 to November 2002.
Observations at C-band (6~cm, 5~GHz) were carried out with the
A-array, observations at X-band (3.6~cm, 8~GHz) were carried out with
both the A- and C-array, and observations at K-band (1.3~cm, 23~GHz) were carried out
with the C-array.  The sources 3C48 and 3C286 were used as flux density
calibrators, and we estimate that the absolute flux density calibration is
$\lesssim 5$\%, based on the scatter in the VLA Flux Density Calibrator
database.  The high-frequency observations at K-band utilized
fast-switching to a nearby phase calibrator, dispaced by $\sim$~5\degr,
in order to mitigate the effect of the atmosphere.

These data were reduced and calibrated using the Astronomical Image
Processing System (AIPS).  In order to better match the largest
spatial scale to which the observations at each wavelength is
sensitive, antenna spacings of $< 20$~k$\lambda$ were not
used. Additionally, in order to achieve synthesized beams of
approximately the same size at each wavelength, the three data sets
were imaged with different weighting of the uv-coverage, ranging from
purely natural weighting at 6~cm to slightly uniform weighting at
1.3~cm.  The data were convolved to identical beams of 1\farcs3 $\times$
1\farcs3 in the imaging process to better facilitate comparison between
wavelengths.  Despite our best efforts, it should be noted that it is
not possible to precisely match the uv-coverage at each wavelength, and
there are slight variations in sensitivity to different spatial scales.
After imaging the data, a primary beam correction was applied
at each wavelength.  The resulting parameters are listed in
Table~\ref{vlaobs}.

\begin{figure*}
\begin{center}
\includegraphics{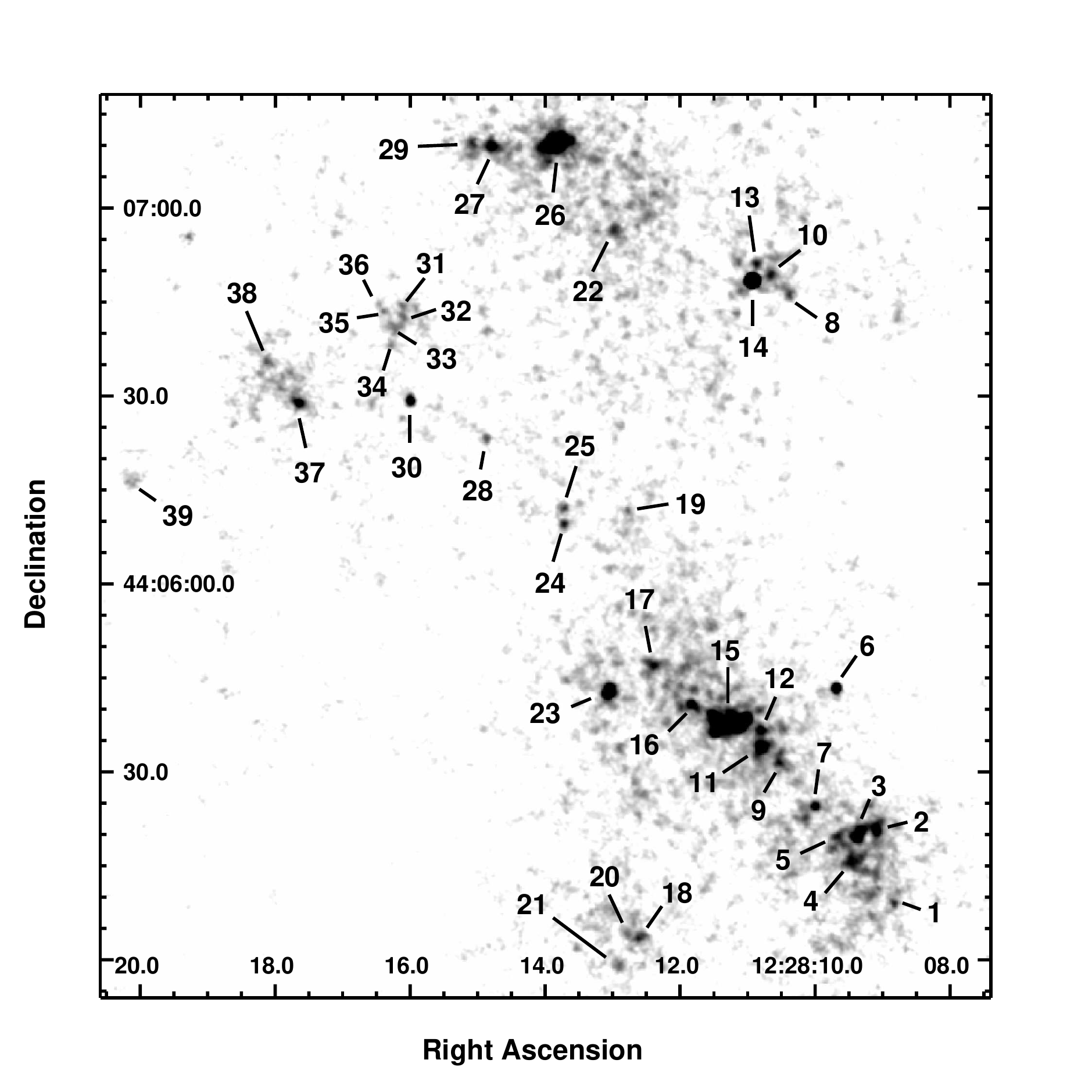}
\caption{VLA 3.6 cm image of NGC~4449.  The radio sources have been labeled
in order of increasing right ascension.  This image has a resolution of 1\farcs3,
a sensitivity of $\sim$~12~$\mu$Jy (RMS), and has been corrected for the primary beam.
\label{sourcemap}}
\end{center}
\end{figure*}

\begin{figure*}
\begin{center}
\includegraphics[scale=0.8]{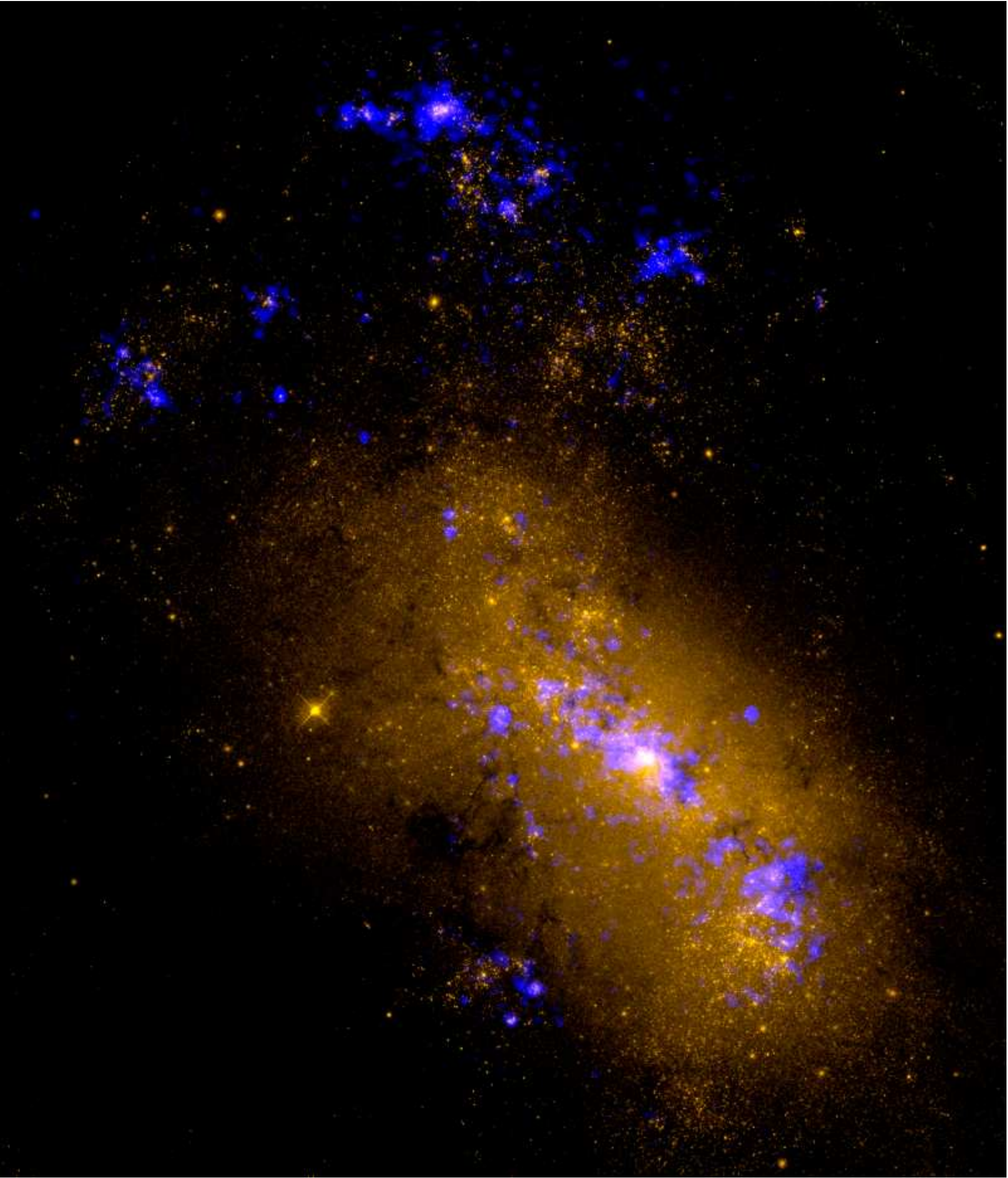}
\caption{An {\it HST}/ACS F550M image (yellow)
overlaid with our VLA 3.6 cm image (blue) highlighting sites of
current star formation in NGC~4449.  The field of view of this
image is $\sim 2\farcm6 \times 3\farcm0$.\label{XVcolor}}
\end{center}
\end{figure*}

We identify 39 compact radio sources towards NGC~4449 which are shown in
Figure~\ref{sourcemap} and labeled in order of increasing right ascension.
Source identification is based on a minimum $3\sigma$ detection (locally) in the
3.6 cm image, which has the highest sensitivity of the three radio images.
Figure~\ref{XVcolor} shows an archival {\it HST} F550M ($\sim V$-band) image
in yellow overlaid with our VLA 3.6 cm image in blue.

\subsection{Ultraviolet, Optical and Near-infrared Imaging with {\it HST}}

A wealth of data on NGC~4449 is contained in the {\it HST}\footnote{Based on observations
made with the NASA/ESA Hubble Space Telescope, obtained [from the Data Archive] at the Space Telescope
Science Institute, which is operated by the Association of Universities for
Research in Astronomy, Inc., under NASA contract NAS 5-26555. These
observations are associated with program \#6716, 7919, 10522 \& 10585.}
archive and we have obtained broad and narrow-band images of the galaxy spanning ultraviolet to
near-infrared wavelengths.  These observations were taken over a period of several
years with the Advanced Camera for Surveys (ACS), the Wide Field and Planetary Camera 2 (WFPC2),
and the Near Infrared Camera and Multiobject Spectrometer
(NICMOS).  The {\it HST} observations have spatial resolutions ranging from
$\sim$~0\farcs1--0\farcs2.  A summary of the {\it HST} observations used in this project
is given in Table~\ref{hstobs}.

The ACS Wide Field Channel (WFC) has a field of view of 202\arcsec $\times$
202\arcsec~and a plate scale of $\sim$~0\farcs05~pixel$^{-1}$.
The ACS pipeline employs the MultiDrizzle \citep{Koekemoer02} software to produce
cosmic ray cleaned, combined images and we have obtained
these data on NGC~4449 through the F435W, F550M, F814W, F658N, and F660N filters.
These observations provide coverage of all of
the locations of the radio sources in NGC~4449 identified in Figure~\ref{sourcemap}.

WFPC2 is composed of four CCDs: three Wide Field (WF) chips and the Planetary Camera (PC).
The plate scales of the PC and WF chips are $\sim$~0\farcs05 pixel$^{-1}$ and $\sim$~0\farcs1
pixel$^{-1}$, respectively.  The WF chips provide an L-shaped field of view of 150\arcsec
$\times$ 150\arcsec~and the PC fills in another 35\arcsec $\times$ 35\arcsec.  All but two
of the radio sources (6 \& 29) in NGC~4449 have WFPC2 coverage through the F170W and F336W filters.
Two exposures were taken through each of these filters
and we combined the calibrated images, chip by chip, and rejected cosmic rays.

\begin{deluxetable*}{cccccc}[!t]
\tabletypesize{\footnotesize}
\tablecolumns{6} 
\tablewidth{0pt} 
\tablecaption{Archival {\it HST} Observations of NGC~4449\label{hstobs}} 
\tablehead{ 
\colhead{Filter}  &  \colhead{Instrument} & \colhead{Description} &
\colhead{Date Observed} & \colhead{Proposal ID} & \colhead{PI}}
\startdata
\cutinhead{Broad and Medium-band Filters}
F170W & WFPC2   & UV                             & 1997 Jul 28, 1998 Jan 9  & 6716 & T. Stecher \\
F336W & WFPC2   & WFPC2 U                        & 1997 Jul 28, 1998 Jan 9  & 6716 & T. Stecher \\
F435W & ACS/WFC & Johnson B                      & 2005 Nov 10-11 & 10585 & A. Aloisi         \\
F550M & ACS/WFC & Narrow V                       & 2005 Nov 18   & 10522 & D. Calzetti          \\
F814W & ACS/WFC & Broad I                        & 2005 Nov 10-11 & 10585 & A. Aloisi         \\
F160W & NIC 3   & H                              & 1998 June 13    & 7919 & W. Sparks        \\         
\cutinhead{Narrow-band Filters}
F658N & ACS/WFC & H$\alpha$ + [NII]$\lambda6584$ & 2005 Nov 17   & 10585 & A. Aloisi          \\
F660N & ACS/WFC & [NII]$\lambda6584$             & 2005 Nov 18   & 10522 & D. Calzetti         \\
\enddata
\end{deluxetable*} 

A snapshot NICMOS observation through the F160W filter was obtained with the NIC~3 Camera
and we have obtained this calibrated image from the archive.  This image, with a field of
view of 51\arcsec $\times$ 51\arcsec~and plate scale of $\sim$~0\farcs2 pixel$^{-1}$,
only covers 7 radio sources (15, 16, 17, 19, 23, 24 \& 25) in the nuclear region of the galaxy.

We register all of the {\it HST} images to match the astrometry of the VLA images, which
is accurate to $\sim$~0\farcs1.  Coincident sources (typically 5--10) are found in pairs of
images and the mean pixel offsets are found between the two images.  
We begin by registering the F658N (H$\alpha$) image to the 3.6 cm radio continuum image
since both observations trace \HII\ regions.  The other {\it HST} images can then be registered
with respect to this or subsequent registered images.
The {\it HST} images are aligned with respect to the
VLA images to a precision of $\sim$~0\farcs2.

\subsection{Infrared Imaging with {\it Spitzer}}

We have obtained infrared images (3.6--24~$\mu$m) of NGC~4449 from the {\it Spitzer Space
Telescope} archive.  We primarily use these data in this paper to qualitatively investigate the
infrared properties of NGC~4449 since these images suffer from significantly lower resolution
($\sim$~3--6\arcsec)
than the VLA and {\it HST} images.  However, despite their limitations, the infrared data are
informative, providing the general IR morphology of NGC~4449 and highlighting regions
associated with warm dust.

\subsection{A Multi-wavelength View of NGC~4449}\label{morph}

\begin{figure*}
\begin{center}$
\begin{array}{cccccc}
\fbox{\includegraphics[width=2.5in]{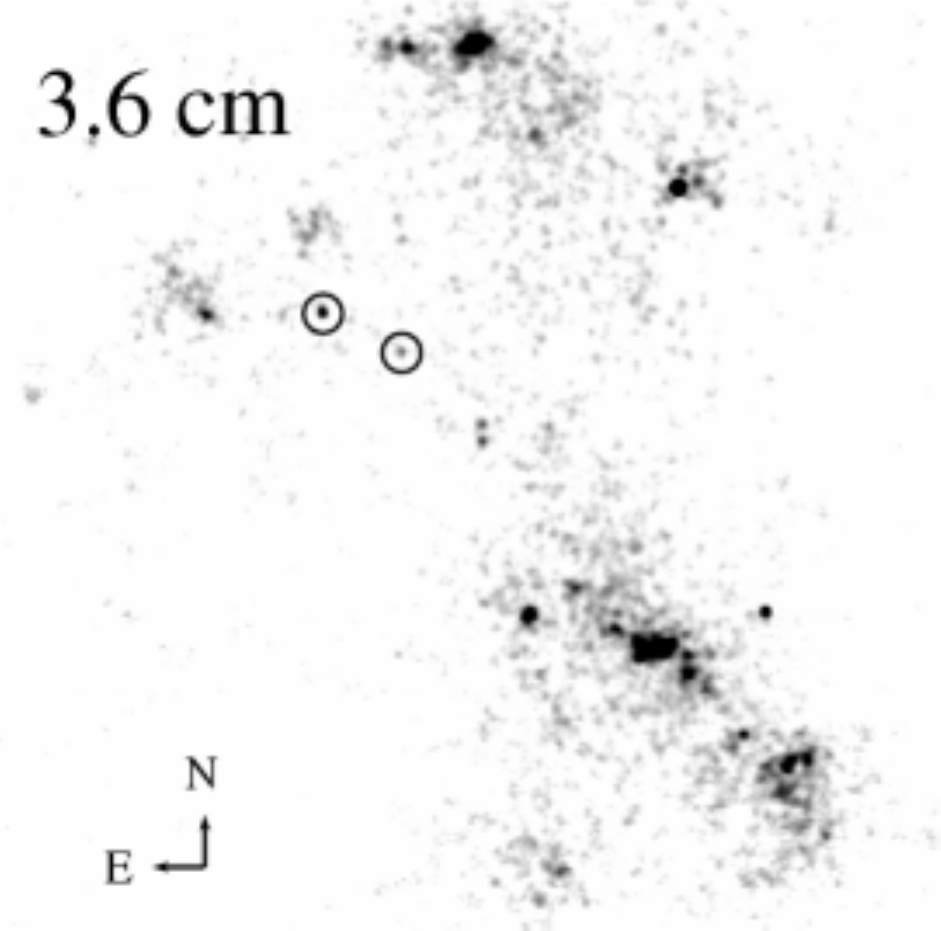}} &
\fbox{\includegraphics[width=2.5in]{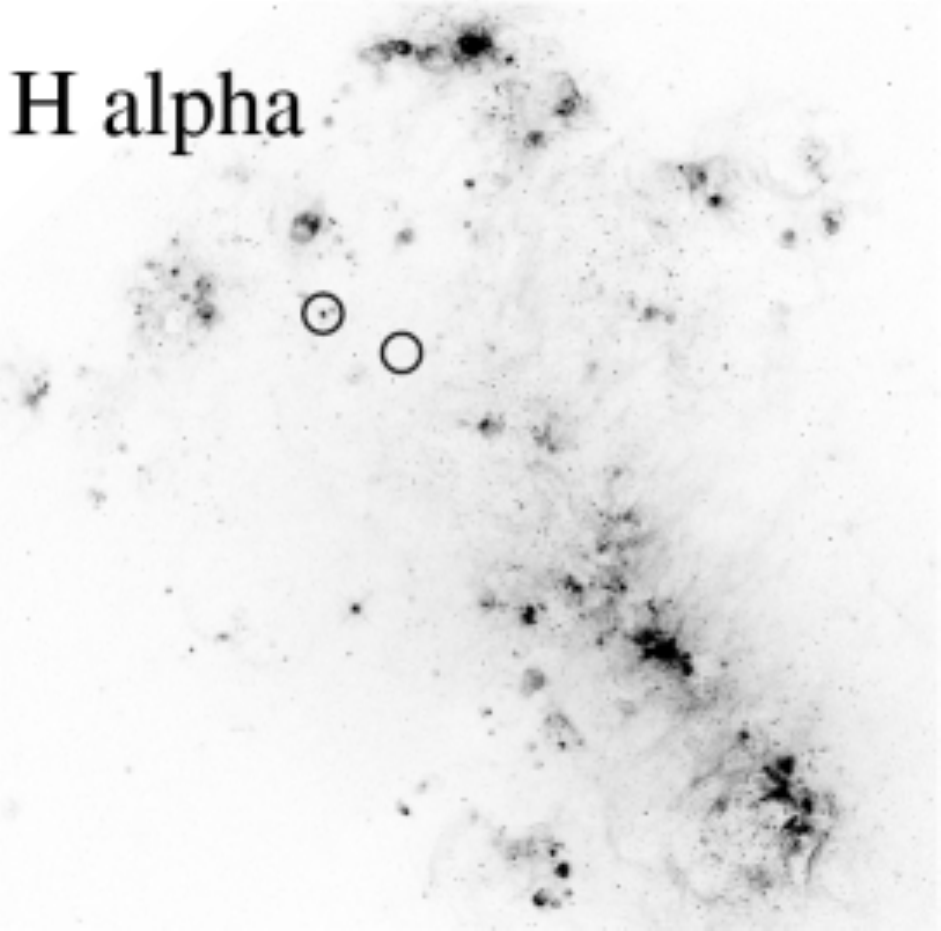}} \\
\fbox{\includegraphics[width=2.5in]{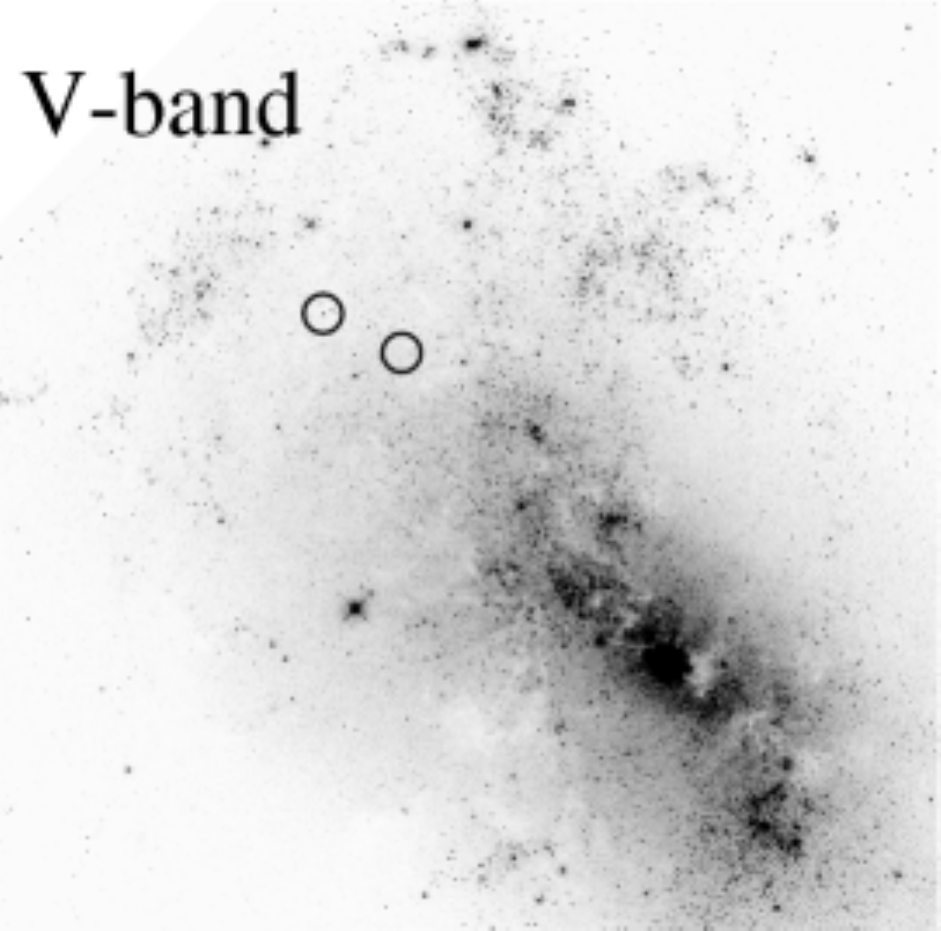}} &
\fbox{\includegraphics[width=2.5in]{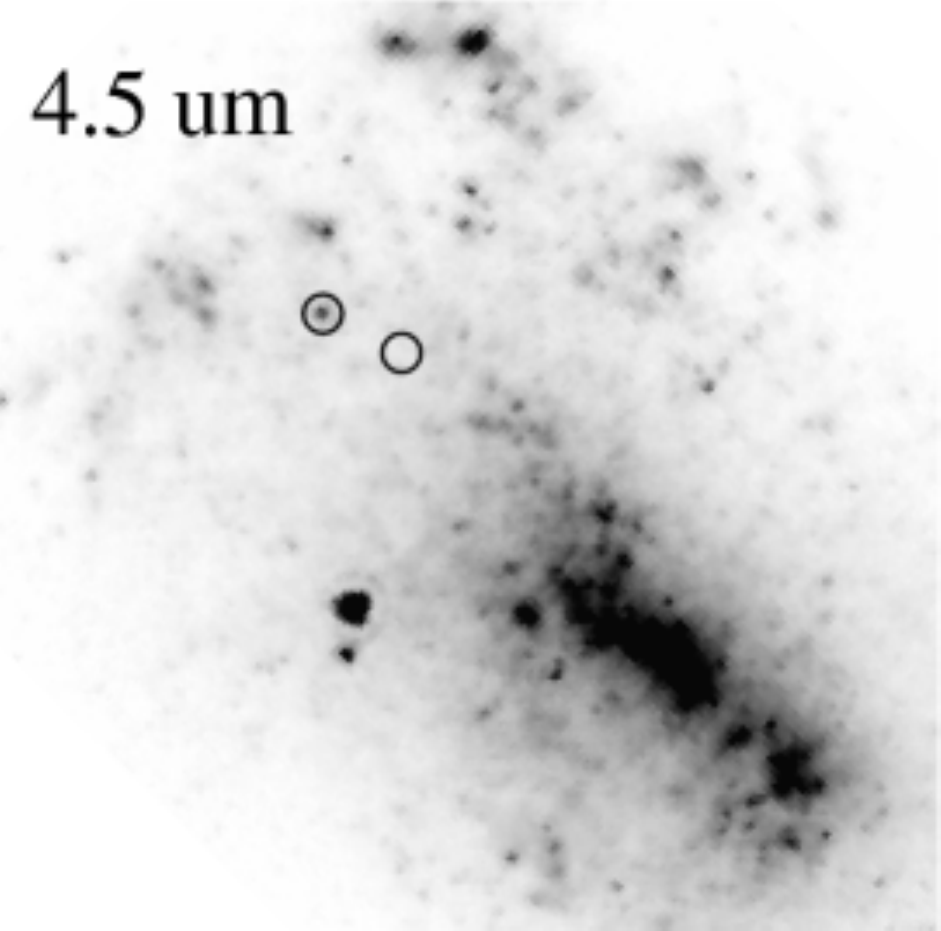}} \\
\fbox{\includegraphics[width=2.5in]{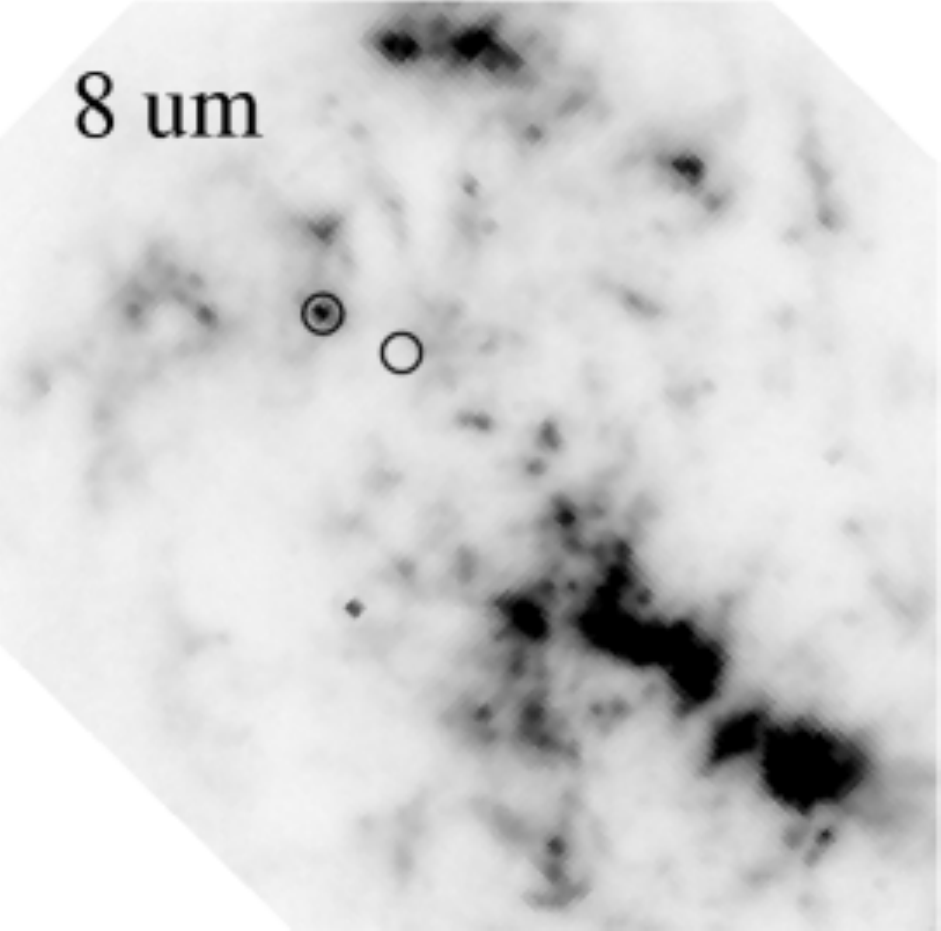}} &
\fbox{\includegraphics[width=2.5in]{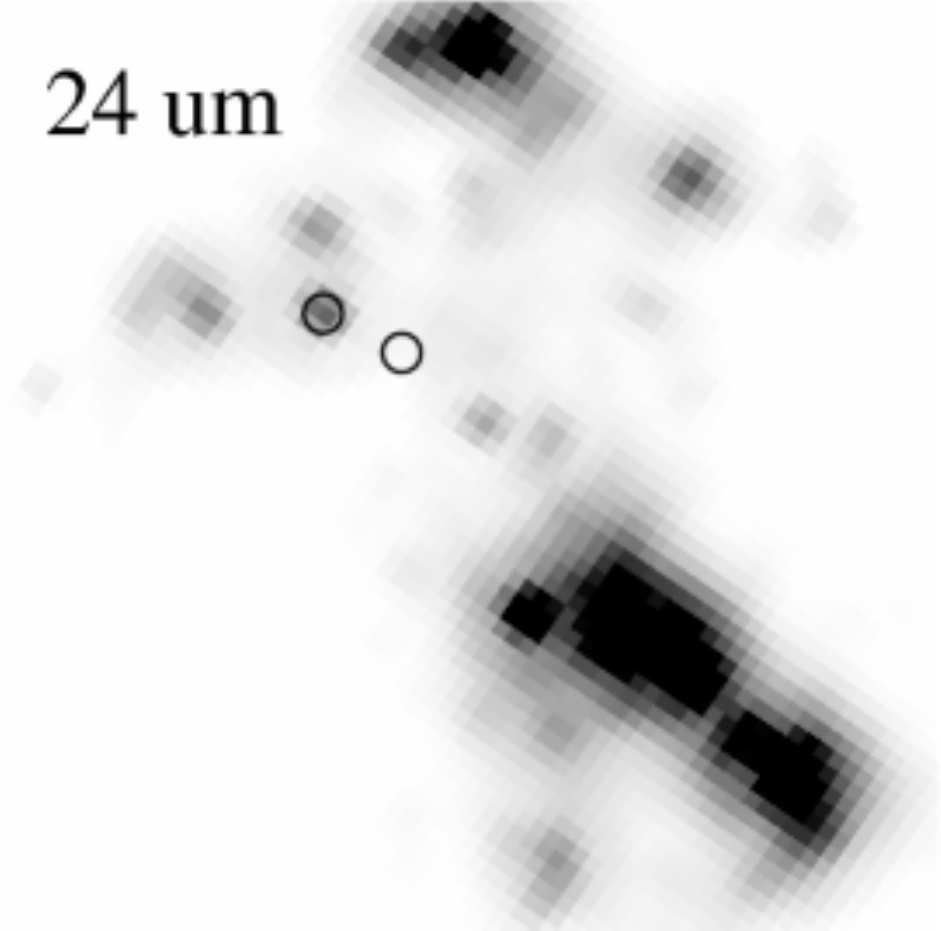}}
\end{array}$
\end{center}
\caption{Multi-wavelength images of NGC~4449 (FOV $\sim 2\farcm5 \times 2\farcm5$).
The following observations are displayed: 3.6~cm (radio continuum), F658N
(H$\alpha$), F550M ({\it V}-band), 4.5~$\mu$m (stellar continuum \& warm dust),
8~$\mu$m (warm dust \& polycyclic aromatic hydrocarbons (PAHs)) and 24~$\mu$m (warm dust).
The circles ($r=3$\arcsec) surround two radio sources:
Source 30, to the NE, has counterparts in all of the other wavebands 
and has therefore already begun to emerge from its gaseous and dusty birth cocoon.
Source 28, on the other hand, does not have counterparts in any of the other images.  The nature
of this source is unclear.
\label{multiwave_ims}}
\end{figure*}

\begin{figure}
\begin{center}$
\begin{array}{ccc}
\includegraphics[width=2.9in]{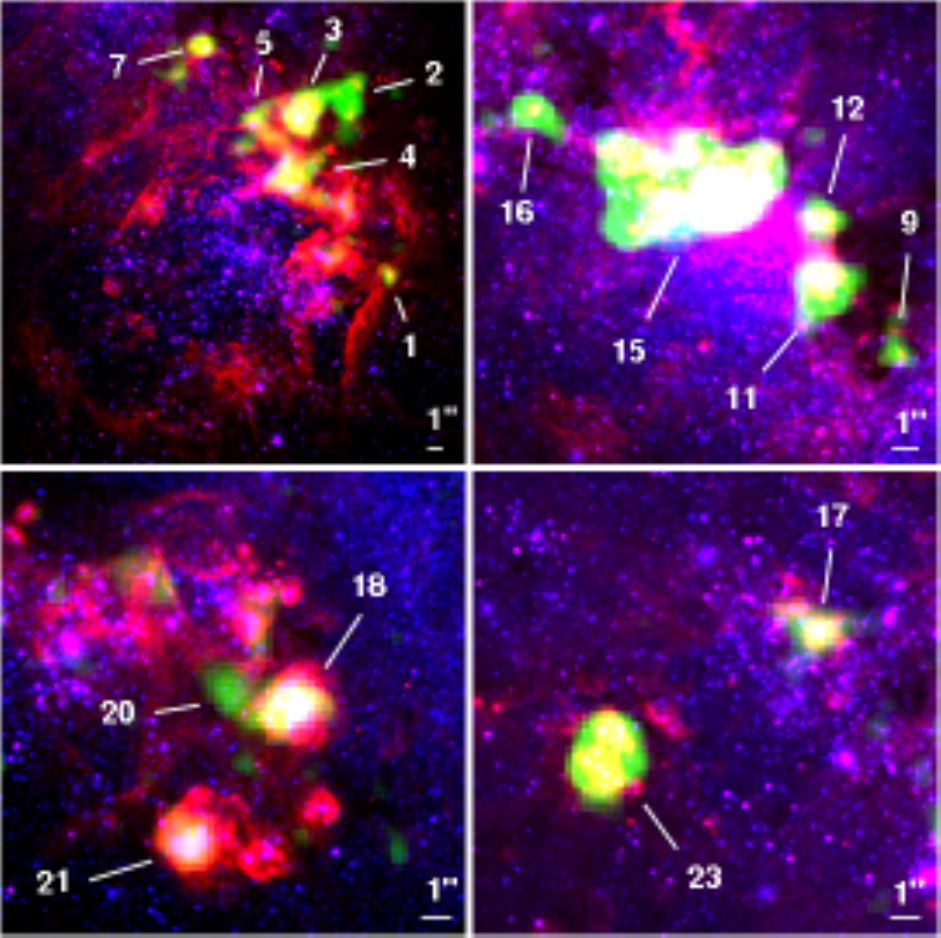} \\
\includegraphics[width=2.9in]{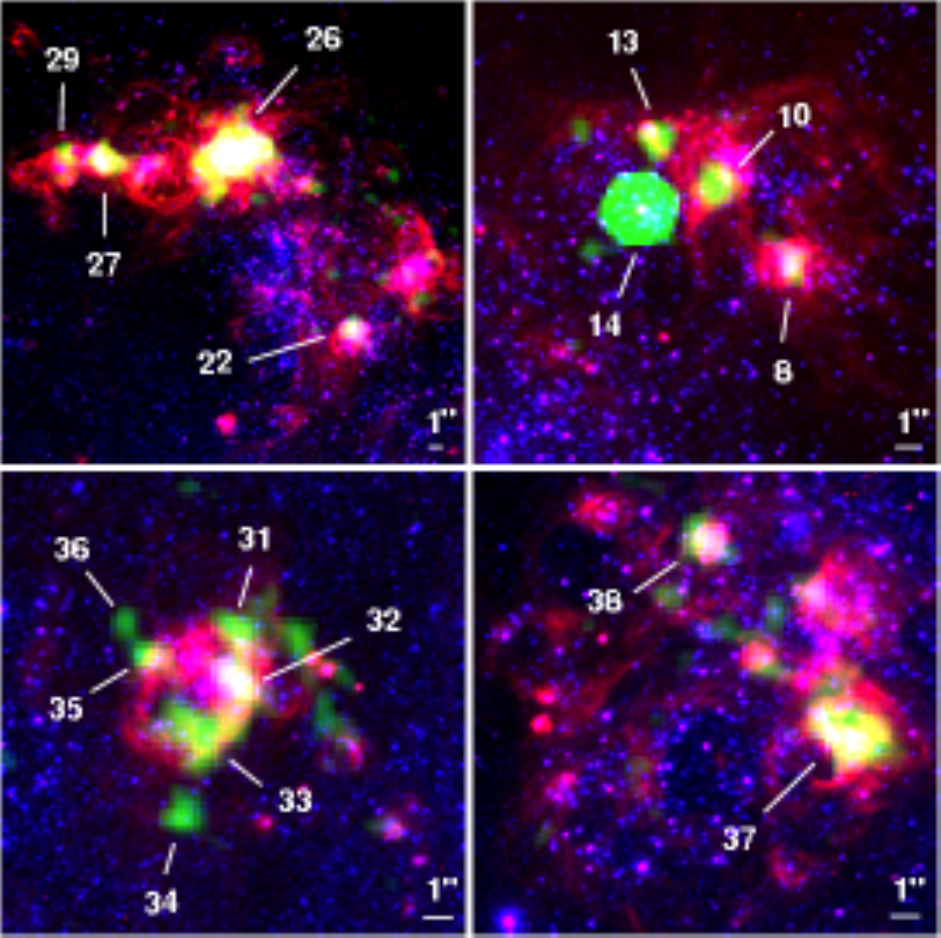} \\
\includegraphics[width=2.9in]{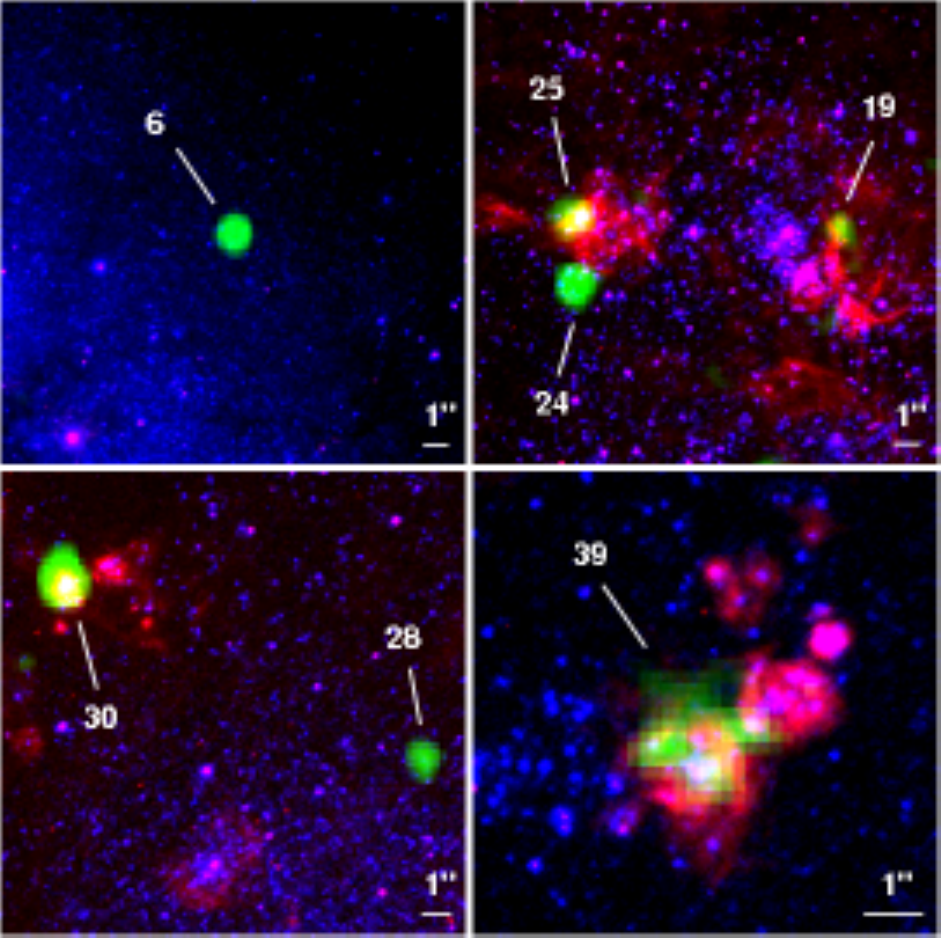} \\
\end{array}$
\end{center}
\caption{Close-up 3-color images of radio-detected sources in NGC~4449: 3.6 cm (green), H$\alpha$ (red),
{\it V}-band (blue).  At the distance of this galaxy (3.9 Mpc), 1\arcsec = 19 pc.
\label{closeups}}
\end{figure}

The high quality data used in this study of embedded massive clusters allows for
a detailed inspection of the starburst morphology of NGC~4449.  Figure~\ref{multiwave_ims}
shows a selection of images of the galaxy at multiple wavelengths.
``Close-up'' 3-color images (3.6~cm, H$\alpha$ and {\it V}-band) are shown in
Figure~\ref{closeups} with the sources labeled according
to Figure~\ref{sourcemap}.   

The similarities between the 3.6~cm and H$\alpha$ observations are immediately apparent in
Figure~\ref{multiwave_ims}.  This likeness is not surprising since both observations trace
dense ionized gas.  
The 3.6~cm and 24~$\mu$m images are also strikingly similar, indicating that the presence
of warm dust is almost exclusively associated with the
ultra-dense \HII\ regions in the galaxy.  

The circles in Figure~\ref{multiwave_ims} are centered on two example radio sources:
one with and one without counterparts in the other wavebands.  Source 30 (to the
NE) has both optical and infrared counterparts, suggesting this stellar cluster
is in the process of emerging from its birth material, all the while ionizing the surrounding gas and
heating the dust.  Source 28 on the other hand, does not have detectable
counterparts in any of the other wavebands.  We will return to the radio sources without
optical or infrared counterparts in \S\ref{pureradio}.  

\section{Multi-wavelength Flux Measurements of the Radio Sources}

\subsection{Methodology}\label{method}

\begin{figure}
\begin{center}
\includegraphics[scale=0.8]{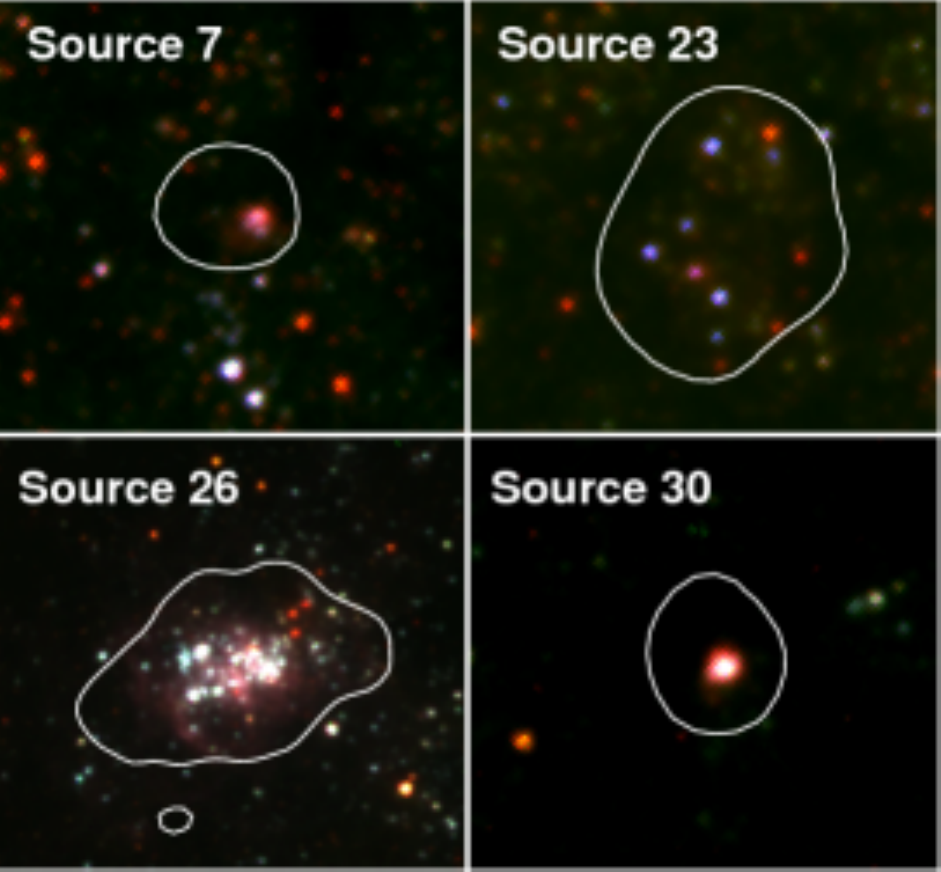}
\caption{Examples of radio-detected star clusters in NGC~4449.  {\it HST}/ACS 3-color
images {\it ($\sim$B, $\sim$V, $\sim$I)} are shown with 3$\sigma$ radio (3.6~cm) contours overlaid. 
\label{clusters_contours}}
\end{center}
\end{figure}

In order to study the emergence of young massive star clusters, we first use radio continuum
imaging to identify potential clusters surrounded by ultra-dense \HII\ regions
(see Fig.~\ref{clusters_contours} for some examples).
We then measure the ultraviolet to radio flux densities of the radio-selected sources
(in the VLA and {\it HST} images) using a new 
photometry code that we developed (SURPHOT), allowing for consistent {\it irregular} apertures and background
annuli across multiple wavebands.  SURPHOT is first used to detect a specified contour level in a selected region of
the 3.6 cm image which is used as a reference.  A source is selected and the contour may be
used for the aperture in all of the images.  
Alternatively, an ellipse, circle or free-form aperture can be used if
appropriate (e.g. nearby yet distinct sources that are contained within the same contour).
A background annulus is constructed by radially expanding the chosen aperture by
specified factors (typically 1.75$\times$ and 2.5$\times$).  The pixel coordinates of the aperture
and background annulus are transformed between images using their FITS World Coordinate Systems (WCS)
which relate image pixels to sky coordinates.  
In this way, flux densities are computed in all (11) of the VLA
and {\it HST} images using identical apertures and background annuli.  (A more detailed description of
SURPHOT is provided in the Appendix.)  The resulting flux densities are used in conjunction with
stellar population synthesis models to derive physical
properties such as extinctions, ages, and masses.  

Our choice to use radio-selected apertures is both scientifically and practically motivated.
The focus of this work is on extremely young massive clusters and centimeter radio emission
is a good indicator of youth.  In addition, the resolution of the VLA images (1\farcs3) is
larger than the resolutions of the {\it HST} images ($\sim$~0\farcs1--0\farcs2).  We note that
using a larger aperture than the intrinsic size of a stellar cluster 
(in the {\it HST} data) allows for the possibility of contamination from an underlying or
surrounding stellar population within the apertures.  However, this effect is not
likely to be very significant since the stellar light is domiated by the clusters of interest
and we perform background subraction in the photometry.

\subsection{The Radio Sources}\label{radiosources}

Thermal radio sources are recognized as having flat or positive radio spectral
energy distributions ($S_\nu \sim \nu^\alpha, \alpha \gtrsim 0$), indicative
of free-free emission from dense ionized gas associated with young massive star
clusters.  Non-thermal sources have decreasing radio SEDs characteristic of
supernova remnants and (background) active galactic nuclei.  Here we are concerned with the
thermal sources in NGC~4449, which must be disentangled from the non-thermal
sources by calculating the spectral indices, $\alpha_{3.6\rm{cm}}^{6.0\rm{cm}}$
and $\alpha_{1.3\rm{cm}}^{3.6\rm{cm}}$, using the measured flux densities.

We use the 3.6 cm radio image as our reference for both source identification
and as the reference image in SURPHOT.  The 1.3 cm image is somewhat less likely
to suffer from non-thermal contamination or self-absorption but the sensitivity is
lower than the 3.6 cm image by a factor of $\sim$~2 at best (the image quality
is degraded farther from the center of the 1.3~cm image).  Source selection is
based on a $3\sigma$ detection threshold, where $\sigma$ is the local RMS noise
(see Fig.~\ref{sourcemap}).  We estimate the minimum uncertainty in the measured radio
flux densities as this local RMS noise.  In many cases,
however, the dominant source of uncertainty comes from the choice of aperture.  For faint sources,
estimates of the background level can also lead to large uncertainties.  We therefore make
multiple flux measurements for each source, varying these parameters.  The fluctuations in these
measurements are accounted for in our final reported values of flux densities and spectral
indices, which are given in Table~\ref{vlaflux} along with the coordinates of the 39
radio sources identified in NGC~4449.
In the cases of weak or non-detections in the 6.0 and 1.3 cm images, $3\sigma$ detection limits
are given for the flux densities.
Corresponding limits are given for the spectral indices.
Sources are designated as either (likely) thermal, non-thermal, or mixed/uncertain 
based on their spectral indices as described above.  Of the 39 radio sources,
13 are thermal, 22 are uncertain or composed of mixed sources and 4 are non-thermal.

Of the 4 definitive non-thermal sources, 3 appear to be associated with the galaxy (Sources 13, 14, \& 39)
while Source 6 is most likely a background AGN with no optical counterpart.  Source 14 is
a well-studied, young (60-200 years) luminous SNR (J1228+441) \citep[see][and references therein]{Lacey07}.
\citet{Lacey07} have given a modern summary of the 30 year history of the radio light curve of J1228+441.
Source 13 may also be a SNR, although it
appears to be at least partially embedded in an \HII\ region.
Source 39 is quite intriguing.  The radio emission is partially associated with an obvious point-like
optical counterpart (probably a SNR) within a larger bi-conical shaped \HII\ region (see
Fig.~\ref{closeups}).  The non-thermal sources within NGC~4449 are interesting in their
own right, but they are excluded from further analysis in this work.

\begin{deluxetable*}{cccrrrrrc}
\tabletypesize{\footnotesize}
\tablecolumns{9} 
\tablewidth{0pt} 
\setlength{\tabcolsep}{0.04in}
\tablecaption{Radio Sources in NGC~4449: Coordinates, Flux Densities and Spectral Indices \label{vlaflux}} 
\tablehead{ 
\colhead{Source}  &  \colhead{R.A.} & \colhead{Decl.} & 
\colhead{$S_{6.0\rm{cm}}$} & \colhead{$S_{3.6\rm{cm}}$} & \colhead{$S_{1.3\rm{cm}}$} &
\colhead{$\alpha_{3.6\rm{cm}}^{6.0\rm{cm}}$} & \colhead{$\alpha_{1.3\rm{cm}}^{3.6\rm{cm}}$} & \colhead{Designation} \\
\colhead{} & \colhead{(J2000.0)} & \colhead{(J2000.0)} & 
\colhead{($\mu$Jy)} & \colhead{($\mu$Jy)} & \colhead{($\mu$Jy)} & \colhead{} & \colhead{} & \colhead{}}
\startdata 
1 & 12 28 08.86 & +44 05 09.0 & 
$<$ 90 & 80(20) & $<$ 210 & $>$ -0.3 & $<$ 1.1 & uncertain \\
 
2 & 12 28 09.12 & +44 05 20.9 & 
$<$ 110 & 120(20) & $<$ 150 & $>$ 0.2 & $<$ 0.3 & likely thermal \\
 
3 & 12 28 09.37 & +44 05 20.2 & 
330(40) & 280(30) & 280(50) & -0.3(0.3) & 0.0(0.2) & likely thermal \\
 
4 & 12 28 09.44 & +44 05 16.3 & 
230(40) & 270(30) & 330(60) & 0.3(0.4) & 0.2(0.2) & thermal \\
 
5 & 12 28 09.70 & +44 05 19.8 & 
$<$ 110 & 130(30) & $<$ 140 & $>$ 0.3 & $<$ 0.1 & likely thermal \\
 
6 & 12 28 09.70 & +44 05 43.5 & 
410(30) & 250(10) & 140(30) & -0.9(0.2) & -0.6(0.2) & non-thermal \\
 
7 & 12 28 10.01 & +44 05 24.7 & 
$<$ 120 & 130(20) & 170(40) & $>$ 0.2 & 0.3(0.3) & thermal \\
 
8 & 12 28 10.38 & +44 06 46.1 & 
$<$ 140 & 150(20) & $<$ 120 & $>$ 0.2 & $<$ -0.2 & uncertain \\
 
9 & 12 28 10.56 & +44 05 31.5 & 
$<$ 120 & 70(20) & $<$ 110 & $>$ -0.8 & $<$ 0.4 & uncertain \\
 
10 & 12 28 10.66 & +44 06 49.3 & 
$<$ 140 & 160(20) & $<$ 130 & $>$ 0.3 & $<$ -0.2 & uncertain \\
 
11 & 12 28 10.81 & +44 05 34.1 & 
160(40) & 230(30) & 220(40) & 0.6(0.5) & 0.0(0.2) & thermal \\
 
12 & 12 28 10.82 & +44 05 36.8 & 
210(50) & 130(20) & 150(40) & -0.8(0.5) & 0.1(0.3) & mixed \\
 
13 & 12 28 10.87 & +44 06 51.5 & 
360(50) & 110(20) & $<$ 140 & -2.1(0.4) & $<$ 0.2 & non-thermal \\
 
14 & 12 28 10.94 & +44 06 48.5 & 
5,920(100) & 2,120(30) & 900(50) & -1.8(0.0) & -0.9(0.1) & non-thermal\tablenotemark{a} \\
 
15 & 12 28 11.30 & +44 05 38.1 & 
2,270(60) & 2,250(60) & 1,510(40) & 0.0(0.1) & -0.4(0.0) & mixed \\
 
16 & 12 28 11.85 & +44 05 41.0 & 
170(50) & 110(20) & 100(30) & -0.7(0.6) & -0.1(0.4) & mixed \\
 
17 & 12 28 12.40 & +44 05 47.1 & 
$<$ 120 & 110(20) & $<$ 80 & $>$ -0.1 & $<$ -0.3 & uncertain \\
 
18 & 12 28 12.63 & +44 05 03.7 & 
$<$ 90 & 150(10) & $<$ 160 & $>$ 0.8 & $<$ 0.1 & likely thermal \\
 
19 & 12 28 12.79 & +44 06 12.0 & 
$<$ 110 & 60(10) & $<$ 70 & $>$ -1.0 & $<$ 0.1 & uncertain \\
 
20 & 12 28 12.82 & +44 05 04.4 & 
$<$ 90 & 40(10) & $<$ 150 & $>$ -1.4 & $<$ 1.3 & uncertain \\
 
21 & 12 28 12.91 & +44 04 59.0 & 
$<$ 100 & 80(10) & $<$ 180 & $>$ -0.3 & $<$ 0.8 & uncertain \\
 
22 & 12 28 12.99 & +44 06 56.3 & 
150(40) & 160(20) & 180(50) & 0.1(0.5) & 0.1(0.3) & thermal \\
 
23 & 12 28 13.08 & +44 05 42.7 & 
500(60) & 470(20) & 480(50) & -0.1(0.2) & 0.0(0.1) & thermal \\
 
24 & 12 28 13.74 & +44 06 09.6 & 
$<$ 110 & 100(10) & $<$ 60 & $>$ 0.0 & $<$ -0.5 & uncertain \\
 
25 & 12 28 13.76 & +44 06 12.3 & 
$<$ 110 & 90(10) & $<$ 70 & $>$ -0.3 & $<$ -0.3 & uncertain \\
 
26 & 12 28 13.86 & +44 07 10.4 & 
1,970(40) & 1,800(40) & 1,360(90) & -0.2(0.1) & -0.3(0.1) & mixed \\
 
27 & 12 28 14.83 & +44 07 10.0 & 
290(40) & 300(20) & 340(90) & 0.1(0.3) & 0.1(0.3) & thermal \\
 
28 & 12 28 14.88 & +44 06 23.2 & 
$<$ 110 & 80(10) & $<$ 70 & $>$ -0.5 & $<$ -0.1 & uncertain \\
 
29 & 12 28 15.09 & +44 07 10.5 & 
$<$ 110 & 160(30) & $<$ 240 & $>$ 0.6 & $<$ 0.4 & likely thermal \\
 
30 & 12 28 16.02 & +44 06 29.3 & 
130(80) & 190(20) & 160(50) & 0.7(1.1) & -0.2(0.3) & likely thermal \\
 
31 & 12 28 16.13 & +44 06 42.6 & 
$<$ 100 & 50(10) & $<$ 120 & $>$ -1.2 & $<$ 0.9 & uncertain \\
 
32 & 12 28 16.14 & +44 06 44.5 & 
$<$ 100 & 60(20) & $<$ 120 & $>$ -0.9 & $<$ 0.7 & uncertain \\
 
33 & 12 28 16.26 & +44 06 40.8 & 
$<$ 100 & 120(20) & $<$ 130 & $>$ 0.3 & $<$ 0.0 & likely thermal \\
 
34 & 12 28 16.32 & +44 06 38.3 & 
$<$ 100 & 50(10) & $<$ 120 & $>$ -1.2 & $<$ 0.8 & uncertain \\
 
35 & 12 28 16.41 & +44 06 43.6 & 
$<$ 100 & 50(10) & $<$ 130 & $>$ -1.3 & $<$ 0.9 & uncertain \\
 
36 & 12 28 16.49 & +44 06 44.7 & 
$<$ 100 & 40(10) & $<$ 130 & $>$ -1.5 & $<$ 1.1 & uncertain \\
 
37 & 12 28 17.69 & +44 06 29.1 & 
$<$ 100 & 240(20) & $<$ 150 & $>$ 1.6 & $<$ -0.5 & uncertain \\
 
38 & 12 28 18.13 & +44 06 35.6 & 
$<$ 100 & 90(20) & $<$ 180 & $>$ -0.3 & $<$ 0.7 & uncertain \\
 
39 & 12 28 20.14 & +44 06 16.2 & 
220(30) & 140(20) & $<$ 320 & -0.8(0.4) & $<$ 0.9 & non-thermal
\enddata 
\tablecomments{Units of right ascension are hours, minutes, and
seconds, and units of declination are degrees, arcminutes, and arcseconds.
The uncertainties in the right ascension and declination are 0.01 seconds
and 0.1 arcseconds, respectively.  Source designations are based on the radio
spectral indices as described in the text.}
\tablenotetext{a}{Source 14 is a well-studied supernova remnant in NGC~4449 \citep[e.g.][]{Lacey07}.}
\end{deluxetable*}  

\subsection{Photometry of the Radio Sources in the {\it HST} Data}

\begin{deluxetable*}{crrrrrrrrrr}
\tabletypesize{\footnotesize}
\tablecolumns{9} 
\tablewidth{0pt} 
\setlength{\tabcolsep}{0.04in}
\tablecaption{{\it HST} Vega Magnitudes of the Radio Sources in NGC~4449 \label{hstmag}} 
\tablehead{ 
\colhead{Source} & \colhead{F170W}  & \colhead{F336W} & \colhead{F435W} & \colhead{F550M} &
\colhead{F814W} & \colhead{F160W} & \colhead{F658N} & \colhead{F660N} }
\startdata
1  & $>$18.40  & 19.93(0.35)  & 21.47(0.08)  & 21.37(0.09)  & 20.05(0.05)
 & \nodata  & 16.72(0.02)  & 18.26(0.07)
  \\
2  & $>$18.00  & $>$20.53  & $>$22.66  & $>$22.51  & $>$21.71  & \nodata
   & 18.96(0.19)  & $>$21.29  
 \\
3  & 15.86(0.08)  & 16.57(0.07)  & 17.65(0.17)  & 17.76(0.12)  & 17.18(0.15)
 & \nodata  & 14.30(0.03)  & 15.76(0.04)
 \\
4  & 14.59(0.04)  & 15.28(0.03)  & 16.74(0.02)  & 16.67(0.03)  & 16.64(0.03)
 & \nodata   & 14.15(0.04)  & 15.38(0.04) 
 \\
5  & 16.82(0.13)  & 17.50(0.05)  & 18.71(0.05)  & 18.88(0.05)  & 18.43(0.08)
 & \nodata   & 16.01(0.09)  & 16.96(0.08)
 \\
6  & \nodata  & \nodata  & 21.02(0.31)  & 19.88(0.10)  & 19.92(0.42)  & \nodata
 &  20.26(0.40)  & 19.95(0.10) 
 \\
7  & 16.59(0.27)  & 17.92(0.12)  & 19.42(0.08)  & 19.50(0.03)  & 19.04(0.14)
 & \nodata   & 15.82(0.05)  & 17.40(0.05)
 \\
8  & 16.94(0.11)  & 17.39(0.05)  & 18.95(0.04)  & 19.12(0.07)  & 18.60(0.04)
 & \nodata  & 15.14(0.07)  & 16.93(0.08)
 \\
9  & $>$18.16  & 17.95(0.03)  & 19.34(0.08)  & 19.05(0.08)  & 18.30(0.06)
 & \nodata   & 17.38(0.07)  & 18.06(0.08)
 \\
10  & 17.71(0.20)  & 17.74(0.09)  & 19.28(0.27)  & 18.83(0.04)  & 18.66(0.23)
 & \nodata   & 15.78(0.08)  & 17.35(0.07)
 \\
11  & $>$17.79  & 17.09(0.26)  & 17.73(0.09)  & 18.04(0.26)  & 16.92(0.10)
 & \nodata   & 15.37(0.08)  & 16.55(0.17)
 \\
12  & $>$17.73  & 18.65(3.05)  & 18.23(0.08)  & 17.86(0.27)  & 17.23(0.06)
 & \nodata    & 15.70(0.16)  & 16.78(0.26)
 \\
13  & $>$18.73  & 18.90(0.07)  & 20.38(0.11)  & 19.95(0.19)  & 19.70(0.14)
 & \nodata    & 16.04(0.08)  & 17.63(0.07)
 \\
14  & 16.40(0.15)  & 17.18(0.12)  & 18.31(0.03)  & 18.67(0.15)  & 17.81(0.02)
 & \nodata  & 16.61(0.20)  & 17.51(0.11)
 \\
15  & 13.48(0.08)  & 13.81(0.03)  & 14.64(0.09)  & 14.29(0.05)  & 13.62(0.07)
 & 12.06(0.04)   & 12.70(0.05)  & 13.37(0.04)
 \\
16  & $>$18.09  & 18.35(0.10)  & 19.40(0.16)  & 19.29(0.05)  & 18.26(0.11)
 & 16.89(0.14)   & 16.88(0.03)  & 17.91(0.17)
 \\
17  & 17.54(0.24)  & 17.72(0.07)  & 18.48(0.15)  & 18.31(0.08)  & 17.63(0.14)
 & 16.23(0.10)   & 15.40(0.04)  & 16.85(0.06)
 \\
18  & 16.11(0.04)  & 17.01(0.02)  & 18.65(0.03)  & 18.91(0.01)  & 18.41(0.05)
 & \nodata  & 14.91(0.02)  & 16.77(0.02)
 \\
19  & $>$18.58  & 19.34(0.13)  & 21.00(0.39)  & 20.43(0.25)  & 20.70(0.49)
 & 19.01(0.36)  & 17.23(0.18)  & 18.52(0.15)
 \\
20  & $>$18.60  & 20.56(0.34)  & $>$23.42  & $>$23.23  & $>$22.62  & \nodata
  & 18.60(0.08)  & 19.20(0.27)
 \\
21  & 15.94(0.04)  & 17.03(0.02)  & 18.56(0.02)  & 18.89(0.02)  & 18.47(0.05)
 & \nodata   & 15.22(0.04)  & 16.94(0.03)
 \\
22  & 15.59(0.03)  & 16.49(0.01)  & 18.08(0.02)  & 17.88(0.01)  & 17.91(0.01)
 & \nodata   & 15.19(0.02)  & 16.66(0.01)
 \\
23  & 16.46(0.15)  & 16.54(0.04)  & 18.22(0.06)  & 18.68(0.03)  & 17.57(0.04)
 & 16.00(0.09)   & 14.46(0.05)  & 15.70(0.04)
 \\
24  & 17.90(0.23)  & 18.71(0.15)  & 19.72(0.13)  & 19.38(0.17)  & 18.92(0.21)
 & 17.05(0.21)   & 17.94(0.14)  & 17.95(0.10)
 \\
25  & 16.71(0.10)  & 17.48(0.06)  & 19.25(0.09)  & 19.00(0.10)  & 18.75(0.10)
 & 17.49(0.11)   & 15.75(0.06)  & 17.02(0.05)
 \\
26  & 14.03(0.01)  & 14.78(0.03)  & 16.20(0.00)  & 16.44(0.02)  & 15.84(0.01)
 & \nodata   & 12.65(0.02)  & 14.41(0.02)
 \\
27  & 16.44(0.06)  & 17.04(0.03)  & 18.29(0.08)  & 18.44(0.02)  & 17.87(0.09)
 & \nodata   & 14.70(0.03)  & 16.32(0.04)
 \\
28  & $>$18.84  & $>$21.34  & 22.03(0.07)  & 21.18(0.09)  & 20.51(0.16)
 & \nodata  & 20.92(0.17)  & $>$21.67 
 \\
29  & \nodata  & \nodata  & 19.59(0.43)  & 19.57(0.27)  & 19.00(0.30)  & \nodata
  & 15.60(0.17)  & 17.21(0.19)  
 \\
30  & $>$18.68  & 18.79(0.07)  & 19.97(0.03)  & 20.25(0.05)  & 18.91(0.03)
 & \nodata   & 15.60(0.02)  & 17.50(0.11)
 \\
31  & $>$18.82  & 18.97(0.14)  & 20.40(0.05)  & 20.65(0.25)  & 19.77(0.03)
 & \nodata  & 16.58(0.12)  & 18.03(0.07)
 \\
32  & 16.85(0.05)  & 17.30(0.03)  & 18.39(0.02)  & 18.61(0.02)  & 18.19(0.02)
 & \nodata   & 15.43(0.02)  & 16.98(0.01)
 \\
33  & 18.86(0.38)  & 19.09(0.37)  & 19.75(0.10)  & 20.32(0.33)  & 19.09(0.07)
 & \nodata   & 16.02(0.12)  & 17.68(0.17)
 \\
34  & $>$18.91  & $>$21.19  & 22.33(0.10)  & 21.96(0.21)  & 21.11(0.04)
 & \nodata   & 18.31(0.11)  & 19.40(0.21)
 \\
35  & $>$19.21  & 19.43(0.07)  & 20.82(0.01)  & 20.98(0.16)  & 20.13(0.03)
 & \nodata   & 17.09(0.19)  & 18.53(0.12)
 \\
36  & $>$19.08  & $>$21.36  & 23.16(0.22)  & 22.36(0.21)  & 21.61(0.21)
 & \nodata   & 19.34(0.15)  & 20.12(0.13)
 \\
37  & 16.91(0.14)  & 17.48(0.03)  & 18.98(0.01)  & 18.90(0.03)  & 18.49(0.00)
 & \nodata  & 15.18(0.02)  & 16.96(0.04)
 \\
38  & 17.17(0.14)  & 17.76(0.03)  & 19.18(0.06)  & 19.10(0.08)  & 18.90(0.06)
 & \nodata  & 16.03(0.02)  & 17.47(0.03)
 \\
39  & 17.17(0.11)  & 18.15(0.05)  & 19.66(0.03)  & 19.78(0.05)  & 19.25(0.04)
 & \nodata  & 15.87(0.03)  & 17.87(0.04)
\enddata
\tablecomments{We adopt a distance modulus of $\sim 28$ for NGC~4449.
A three-dot ellipsis indicates no data was available.}
\end{deluxetable*}  

We use the 3.6 cm radio map as the reference image in SURPHOT (see Appendix) to obtain photometry of the
radio sources in the {\it HST} data.  {\it HST} system magnitudes with Vega zeropoints
are reported in Table~\ref{hstmag}.  Error estimates include contributions from Poisson noise
and uncertainties from varying the apertures (see \S\ref{radiosources}).

\section{Properties of the Emerging Star Clusters in NGC~4449}\label{propsec}

\subsection{The Models}\label{models}

In order to estimate the physical properties of the radio selected
sources in NGC~4449, we compare our data to the latest {\scriptsize STARBURST}99 population
synthesis models (Version 5.1) of \citet{Leitherer99}.  We have run simulations
at two metallicities, Z=0.004 and Z=0.008, as appropriate for NGC~4449 \citep{Lequeux79}.  
We adopt an instantaneous burst of $10^4$~M$_\odot$ with a Kroupa IMF, the Geneva evolutionary
tracks with high mass loss, and the Pauldrach/Hillier atmospheres.
Ages, masses, and extinctions are estimated using two distinct methods: 1) comparing
the nebular emission from the \HII\ regions to the model output (\S\ref{nebprop}); and 2) fitting model spectral
energy distributions (SEDs) to the broad-band {\it HST} flux densities (\S\ref{sedprop}).

\subsection{Results from the Nebular Emission}\label{nebprop}

\subsubsection{Ages}\label{ages}

We estimate ages of the radio sources in NGC~4449 by comparing their measured
H$\alpha$ equivalent widths with the model predictions as a function of age.
For an instantaneous burst of star formation, the equivalent width of H$\alpha$ emission
is a reliable age indicator for young clusters since it measures the ratio of the ionizing
flux (from massive stars) to the total continuum flux \citep[e.g.,][]{Alonso-Herrero96}.
This ratio is strongly dependent on age for stellar populations $\lesssim$~10~Myr, so long
as the most massive stars have begun to die, thereby decreasing the ionizing flux.
In addition, if the extinction is the same to the stars and the gas, H$\alpha$ equivalent width
will be independent of extinction.

We calculate H$\alpha$ equivalent widths,
$W_{{\rm H}\alpha}$, using Equations~\ref{haew} \& \ref{haflux} below:

\begin{equation}
W_{{\rm H}\alpha} = {F_{{\rm H}\alpha} \over f_{\rm cont}^{({\rm H}\alpha)}},
\label{haew}
\end{equation}

\noindent
where $F_{{\rm H}\alpha}$ is the total H$\alpha$ flux (erg~s$^{-1}$~cm$^{-2}$) and $f_{\rm cont}^{({\rm H}\alpha)}$
is the continuum flux density (erg~s$^{-1}$~cm$^{-2}$~\AA$^{-1}$) at 6563~\AA.  The F658N filter contains
both H$\alpha$ and [NII]$\lambda6584$ emission so the contribution from the [NII] line must be subtracted.
$F_{{\rm H}\alpha}$ is given by

\begin{eqnarray}
\label{haflux} F_{{\rm H}\alpha} = \left[\left(f_{\rm F658N} - f_{\rm cont}^{({\rm H}\alpha)}\right)
\Delta\lambda_{\rm F658N}\right] - \\
\left[\left(f_{\rm F660N} - f_{\rm cont}^{({\rm NII})}\right) \Delta\lambda_{\rm F660N}\right],
\nonumber
\end{eqnarray}

\noindent
where the first term in square brackets is the total H$\alpha$ + [NII]$\lambda6584$ flux
and the second term in square brackets is the total [NII]$\lambda6584$ flux.
Differences in throughput of the filters at the wavelengths of H$\alpha$ and [NII]
are accounted for in Equation~\ref{haflux}.
The measured flux densities and widths of the filters are given by  
$f_{\rm F658N}$,$f_{\rm F660N}$ and $\Delta\lambda_{\rm F658N}$,$\Delta\lambda_{\rm F660N}$, respectively.
The flux densities of the continuum values, $f_{\rm cont}^{({\rm H} \alpha)}$ and $f_{\rm cont}^{({\rm NII})}$,
are found by interpolating between the F550M and F814W data.  

We compare the calculated H$\alpha$ equivalent widths, $W_{{\rm H}\alpha}$, to the models in order to estimate
ages for the radio sources in NGC~4449.  The thermal radio sources in NGC~4449 have H$\alpha$ equivalent
widths between $\sim$~340 and $\sim$~2460 \AA, implying they have ages $\lesssim$~5~Myr.
However, it should be noted that the {\scriptsize STARBURST}99 models reflect a fully
sampled stellar IMF.  For clusters with masses $\lesssim 10^6$~M$_\odot$, it
becomes increasingly unlikely that a given cluster will contain extremely
massive stars.  As a result, lower mass clusters will tend to produce
less ionizing flux for the same age than would be predicted by simply
scaling the {\scriptsize STARBURST}99 predictions. 
In addition, the possibility of contamination from an underlying or surrounding stellar population
to the continuum measurements as a result of using radio-selected apertures (see \S\ref{method})
could artificially decrease the measured H$\alpha$ equivalent widths.
Therefore, the radio-detected clusters may be even younger than derived here (Table~\ref{physprop}).

\subsubsection{Ionizing Fluxes}

\begin{deluxetable*}{cccccc}[!t] 
\tabletypesize{\footnotesize}
\tablecolumns{6} 
\tablewidth{0pt} 
\tablecaption{Properties of the Thermal Radio Sources \label{physprop}} 
\tablehead{ 
\colhead{Source} & \colhead{Log $W_{{\rm H}\alpha}$} & \colhead{Age} & \colhead{$Q_{\mbox{Lyc}}$} & \colhead{Mass}  & 
\colhead{$A_{\rm V}$} \\
\colhead{} & \colhead{(\AA)} & \colhead{(Myr)}  &  \colhead{($10^{49}$ s$^{-1}$)}  & 
\colhead{($10^3$ M$_\odot$)} & (mag) }
\startdata 
2 & $>$3.10 & $<$3.7 & 17(3) & $<$7 & 4.3(0.3) \\ 
3 & 3.00(0.28) & (1.2)4.0(0.8) & 40(4) & 21(2) & 0.3(0.1) \\
4 & 2.68(0.07) & (0.2)4.9(0.1) & 39(5) & 40(5) & 0.2(0.1) \\
5 & 2.73(0.16) & (0.4)4.8(0.3) & 19(4) & 18(3) & 1.5(0.3) \\
7 & 3.12(0.22) & (0.6)3.3(1.0) & 18(3) & 6(1) & 1.0(0.2) \\
11 & 2.53(0.50) & (2.1)5.2(2.2) & 33(4) & 44(5) & 1.5(0.4) \\
18 & 3.26(0.07) & (0.2)2.9(0.1) & 21(2) & 6(1) & 0.1(0.1) \\
22 & 2.79(0.03) & (0.1)4.6(0.1) & 23(2) & 19(2) & 0.7(0.1) \\
23 & 3.21(0.08) & (0.1)2.9(0.7) & 68(3) & 19(1) & 1.2(0.1) \\
27 & 3.13(0.14) & (0.8)3.6(0.5) & 42(3) & 16(1) & 0.8(0.1) \\
29 & 3.20(0.58) & (2.9)2.9(2.1) & 23(4) & 6(1) & 1.0(0.3) \\
30 & 3.39(0.09) & (1.0)2.6(0.2) & 27(2) & 6(1) & 1.2(0.1) \\
33 & 3.24(0.48) & (2.9)2.9(1.8) & 18(3) & 5(1) & 1.2(0.2) 
\enddata 
\tablecomments{H$\alpha$ equivalent widths, ages, ionizing fluxes, masses
and extinctions of the thermal radio sources in NGC~4449.
These values were derived from the nebular emission from the
\HII\ regions (\S\ref{nebprop}).}
\end{deluxetable*}  

The ionizing flux of a starburst region, $Q_{\rm Lyc}$, can be estimated from
its radio luminosity density following \citet{Condon92}:

\begin{eqnarray}
\label{Qlyc} \left({Q_{\rm Lyc} \over {\rm s^{-1}}}\right) \gtrsim 6.3\times10^{52}
\left({T_e \over 10^4{\rm ~K}}\right)^{-0.45} \left({\nu \over {\rm GHz}}\right)^{0.1} \\
\times \left({L_{\nu, \rm thermal} \over 10^{27} {\rm ~erg ~s^{-1} ~Hz^{-1}}}\right).
\nonumber
\end{eqnarray}

\noindent
Equation \ref{Qlyc} is applicable to thermal radio emission
but provides only a lower limit since some of the ionizing photons may
be absorbed by dust and the radio emission may be optically thick.  
Studies of Lyman continuum extinction in \HII\ regions find that   
the fraction of ionizing photons absorbed by dust can be quite large \citep{Inoue01,Dopita03}, leading
to derived ionizing fluxes (and therefore masses) that can be artificially low by up to a factor
of 10.  Nonetheless, we adopt an electron temperature of $T_e = 10^4$~K and
use the 3.6 cm radio luminosities to calculate $Q_{\rm Lyc}$ using Equation~\ref{Qlyc} for the radio sources
in NGC~4449.  The 1.3 cm radio luminosities are somewhat less likely
to be contaminated by non-thermal emission or be optically thick, but the 1.3 cm image does not
have the requisite sensitivity.  The derived $Q_{\rm Lyc}$ of the thermal radio sources range
from $\sim$~20--70~$\times 10^{49}$~s$^{-1}$, or equivalently \citep{Vacca96}, $\sim$~20--70 O7.5 V stars.

\subsubsection{Masses}

The ionizing flux and age of a thermal radio source can be used estimate its total stellar mass.
The models predict the number of ionizing photons at a given age for an input total stellar
mass.  Using the age derived from the H$\alpha$ equivalent width, we determine the expected
ionizing flux at that age, given by the model.  We assume that mass scales with ionizing flux
and estimate the total stellar mass by multiplying the input model mass by the ratio of
the observed to expected ionizing flux.  The masses of the thermal radio sources derived
using this method are in the range $\sim$~0.5--5~$\times 10^4$~M$_\odot$.

\subsubsection{Extinctions}\label{extinction}

Global extinction of the ionized nebular gas is estimated by comparing the measured H$\alpha$ flux to
the predicted H$\alpha$ flux (with no extinction) derived from the thermal radio flux density,
$S_{\nu, \rm thermal}$.  The radio emission does not suffer from extinction per se, although again we note
that the UV photons can suffer significant absorption before ionizing the gas, which is not
accounted for in Equation~\ref{Qlyc}.  We follow \citet{Condon92}
to obtain the predicted H$\alpha$ flux:

\begin{eqnarray}
\left({F_{{\rm H}\alpha,{\rm predicted}} \over {\rm erg~s^{-1}~cm^{-2}}}\right) \sim 0.8 \times 10^{-12}
\left({T_e \over 10^4{\rm ~K}}\right)^{-0.59} \\
\times \left({\nu \over {\rm GHz}}\right)^{0.1} \left({S_{\nu, \rm thermal} \over {\rm mJy}}\right).
\nonumber
\end{eqnarray}

\noindent
The extinction of H$\alpha$ is then given by

\begin{equation}
A_{{\rm H}\alpha} = -2.5 \times {\rm log}\left({F_{{\rm H}\alpha,{\rm observed}} \over F_{{\rm H}\alpha,{\rm predicted}}} \right)
\end{equation}

A 30~Doradus extinction curve \citep{Fitzpatrick85,Fitzpatrick90,Misselt99}
is used to convert $A_{{\rm H}\alpha}$ to $A_{\rm V}$, the extinction at 5500~\AA.
The thermal radio sources have measured nebular extinctions of $A_{\rm V} \lesssim 1.5$ with the exception
of Source 2, which is highly obscured and has $A_{\rm V} \approx 4.3$.  
Measured extinctions, H$\alpha$ equivalent widths, ages, ionizing fluxes and masses of the
thermal radio sources in NGC~4449 are listed in Table~\ref{physprop}. 

\subsection{Results from SED Fitting: Evidence for {\it I}- and {\it H}-band Excesses}\label{sedprop}

In addition to using the nebular emission, we also estimate the physical properties of the
radio selected sources in NGC~4449 using the stellar emission by comparing a grid of model
spectral energy distributions to the broad-band {\it HST} flux densities.
The grid consists of models with ages up to 10 Myr in increments of $\Delta t=0.1$~Myr
and extinctions in increments of $\Delta{A}_{\rm V}=0.1$.  We
typically run the grid from $A_{\rm V}$ = 0 to 5 magnitudes which yields a total of 5,151 model SEDs to compare
to each observed SED.  Galactic foreground extinction is also accounted for (NGC~4449 has a Galactic reddening of
$E(B-V) = 0.019$ \citep{Schlegel98}).

\begin{deluxetable}{ccc}[!t]
\tabletypesize{\footnotesize}
\tablecolumns{3} 
\tablewidth{0pt} 
\tablecaption{Results of SED Fitting\label{curves}} 
\tablehead{ 
\colhead{Extinction Curve} & \colhead{Z=0.004} & \colhead{Z=0.008}}
\startdata
\cutinhead{Fitting the {\it UV, U, B} \& {\it V} Data}
30 Doradus  & 0.032(0.028) & 0.032(0.027) \\
Galactic & 0.040(0.034) & 0.040(0.035) \\
Calzetti & 0.043(0.036) & 0.042(0.036) \\
\cutinhead{Fitting the {\it UV, U, B, V} \& {\it I} Data}
30 Doradus  & 0.061(0.041) & 0.061(0.038) \\
Galactic & 0.055(0.038) & 0.054(0.037) \\
Calzetti & 0.058(0.038) & 0.056(0.037) \\
\enddata 
\tablecomments{The goodness-of-fit parameter, $\sigma$, and its standard deviation, averaged over all possible
sources, is listed for 12 runs of SED fitting.  Excluding the {\it I}-band flux and using a 30 Doradus
extinction curve fits the data significantly better than the other runs.  A full discussion is given in the text.}
\end{deluxetable} 

Each model spectral energy distribution, with a given age and extinction, is convolved with the {\it HST}
total throughput curves corresponding to
the broad-band filters listed in Table~\ref{hstobs} before it is compared to the observations.
Once we have obtained the model flux densities for each broad-band filter, we compute
a goodness-of-fit statistic, $\sigma$, given by the standard deviation of the difference between
the log of the observed flux densities and the log of the model flux densities, weighted by the 
measurement errors of the observed flux densities.  In the case where we only have an upper limit for
an observed flux density, we use this limit to exclude models but do not include this point in the fit.
The model that minimizes $\sigma$ is chosen as the best fit.  A mass estimate is obtained by scaling the input
model mass by the weighted mean offset between the observed and best-fit model flux densities
(again done in the log).

We have performed SED fitting on the radio selected sources in NGC~4449 with detections in the {\it HST}
data.  The most heavily obscured sources are not fit since we only have upper limits for all of the flux
densities.  The strongest Balmer lines, H$\alpha$ and H$\beta$ are not contained in any of the broad-band
filters used here, although the F435W filter does include H$\gamma$ and H$\delta$.  Originally, we included up
to five filters in the fitting process: F170W~{\it ($\sim$UV)}, F336W~{\it ($\sim$U)}, F435W~{\it ($\sim$B)},
F550M~{\it ($\sim$V)}, and F814W~{\it ($\sim$I)}.  However, we found that even the best-fitting models did not visibly fit
the data in many cases because of an apparent excess in the {\it I}-band.  To check there were no calibration
problems in the F814W data, we obtained WFPC2 F814W images of NGC~4449 from the archive.
The results were consistent with our findings in the ACS data.  

To investigate further, we ran our SED fitting routine twelve times on each possible source: including and
excluding the F814W data point, using models with metallicities of Z=0.004 and Z=0.008,
and applying three different extinction curves: the Galactic extinction curve \citep{Cardelli89},
the starburst obscuration curve \citep{Calzetti00}, and a 30 Doradus extinction curve
adopted from \citet[][Table~3]{Misselt99} and \citet[][Table~6]{Fitzpatrick85}, using
the parameterization given by \citet{Fitzpatrick90}.  For each of the twelve runs, we have computed the
mean $\sigma$ (goodness-of-fit statistic) of all the sources and its standard deviation (Table~\ref{curves}).
These values measure how well a family of models (metallicity, extinction curve, including/excluding $I$-band data)
fit the data set as a whole.  Table~\ref{curves} shows that excluding the {\it I}-band data and
using the 30 Doradus extinction curve fits the data significantly ($\sim$~20--50\%) better than the
other families of models, but that there is no significant difference between the fits using either metallicity.

It is also evident in Table~\ref{curves} that excluding the {\it I}-band data significantly improves the
fits, regardless of which extinction curve is used.  A better fit is expected if using fewer
data points (i.e., excluding the $I$-band); however, we ran Monte Carlo simulations which showed
that including one more data point in the fits should only increase $\sigma$ by $\sim$~2--5\%, not 33--91\%
as seen in Table~\ref{curves}.  Data was simulated by adding noise (representative of measurement errors)
to model SEDs and then convolving the SEDs with the appropriate throughput curves.  The simulated data were compared
to the original SEDs and the goodness-of-fit statistic, $\sigma$, was calculated with and without the $I$-band data. 

\begin{figure}
\begin{center}
\includegraphics[scale=0.6]{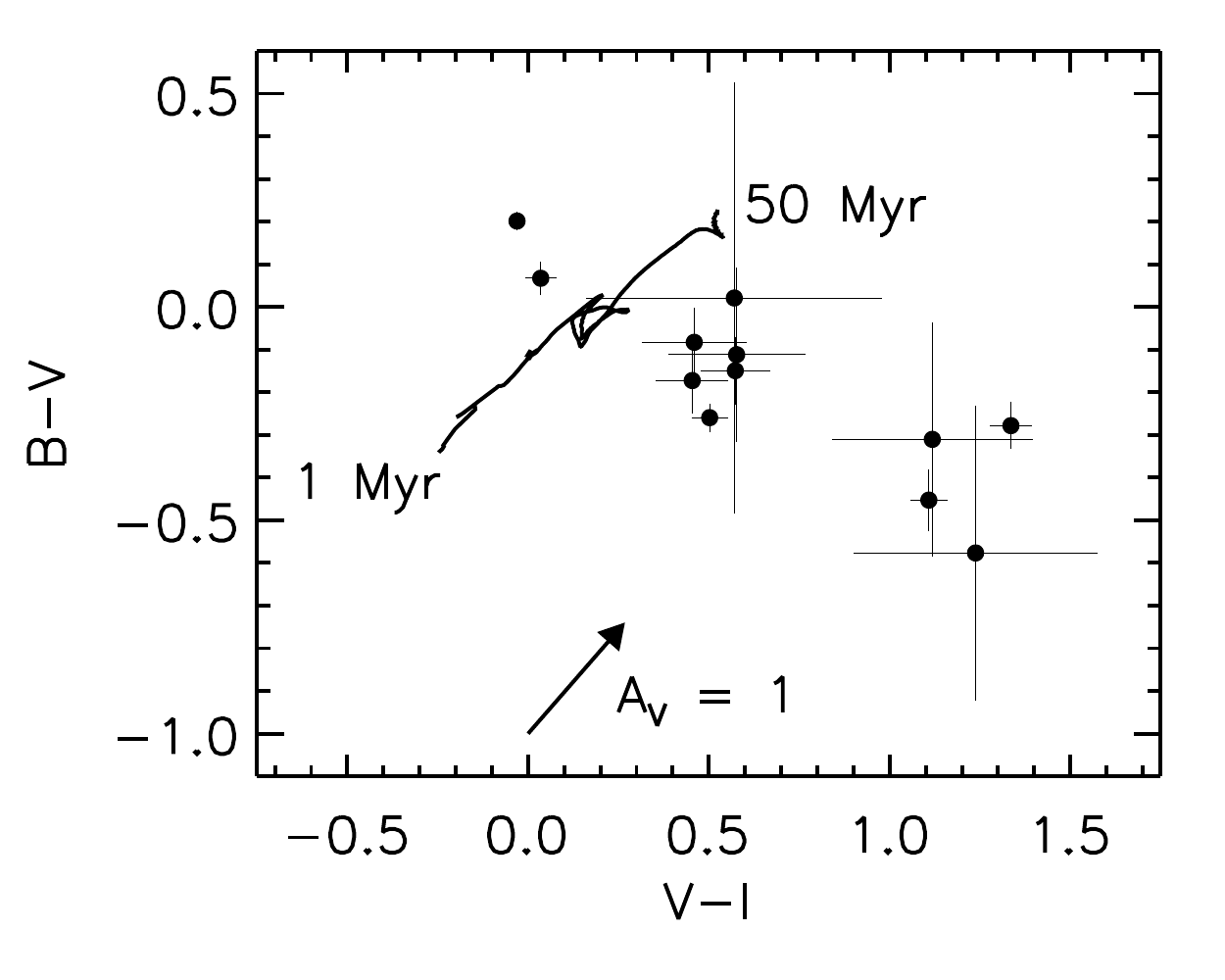}
\caption{Color-color diagram of the thermal radio sources in NGC~4449.  This plot of
$B-V$ (F435W--F550M) versus $V-I$ (F550M--F814W) illustrates that the reddest clusters in
$V-I$ are the bluest in $B-V$, contrary to the expected trend if extinction
were responsible for the red $V-I$ values (i.e. the {\it I}-band excess).
A model evolutionary track and a reddening vector (using the 30 Doradus extinction curve)
are also shown for comparison.
\label{BVvsVI}}
\end{center}
\end{figure}

It should be emphasized that extinction of the stellar continuum cannot account for the observed SEDs
when including the {\it I}-band data.  Figure~\ref{BVvsVI} shows a plot of $B-V$ (F435W--F550M)
versus $V-I$ (F550M--F814W) illustrating
that the reddest clusters in $V-I$ are the bluest in $B-V$, contrary to the expected trend if extinction
were responsible for the red $V-I$ values.  The model evolutionary track and a reddening
vector (using the 30 Doradus extinction curve) are also shown for comparison.
This anomalous color-color diagram, along with our results from SED fitting, leaves little doubt that the
{\it I}-band excess is real.  

\begin{figure*}[!t]
\begin{center}
\includegraphics[scale=.85]{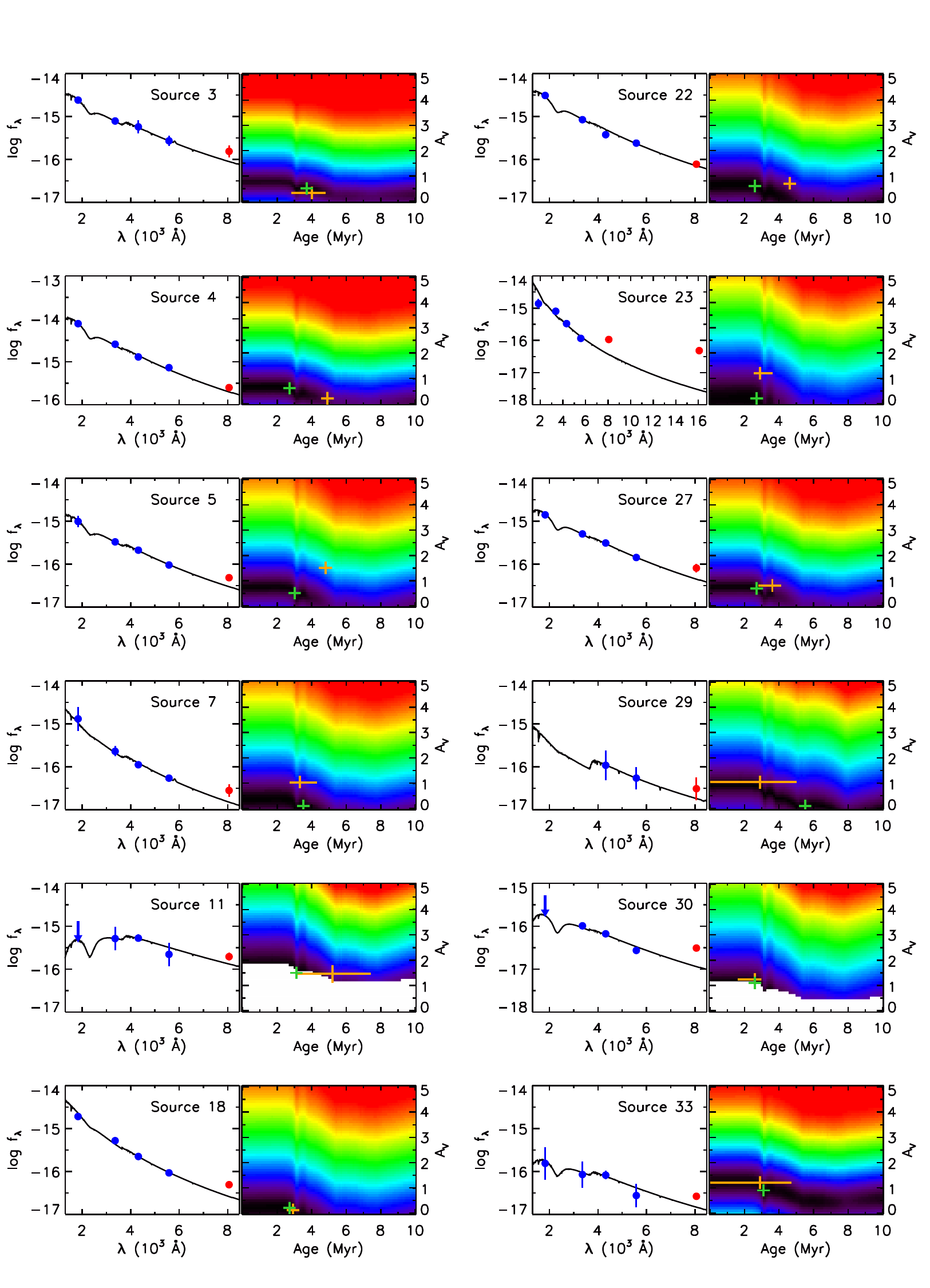}
\caption{UV to near-IR observed flux densities (erg~s$^{-1}$~cm$^{-2}$~\AA$^{-1}$)
and best-fitting model SEDs for the thermal radio sources.
The blue data points ($UV, U, B, V$) are included in the fit but the red points ($I, H$) are not. 
Grids of $\sigma$ (goodness-of-fit parameter) are also shown for models of various ages and extinctions.
The white space in the grids indicate models that were rejected based on the ultraviolet detection limits.
A green cross indicates the best-fitting model and, for comparison,
an orange cross indicates the values obtained from the nebular emission as described in \S\ref{nebprop}.
Source 2 is not included since it was not detected in the broad-band optical filters.
\label{sedthermal}}
\end{center}
\end{figure*}

On average, the {\it I}-band flux densities are $\sim$~60\% higher (0.5 magnitudes brighter) than the
predicted values from the best-fitting model SEDs.  In the most extreme cases, the measured {\it I}-band
flux densities exceed the model values by more than a factor of 2.5.  An even greater excess is present
in the F160W~{\it (H)} filter (also not included in the fits) with the data brighter than
the best-fitting SED by an average of 1.6 magnitudes, or $\sim 4 \times$ the flux.
We will discuss possible origins of the {\it I}- and {\it H}-band excesses in \S\ref{irexcess}.

Figure~\ref{sedthermal} shows the observed flux densities and best-fitting model SEDs
for the thermal sources that have detections in the {\it HST} data.  The models have
a metallicity of Z=0.004, use the 30 Doradus extinction curve, and exclude the {\it I}- and {\it H}-band data
in the fits.  The grids of $\sigma$ are also shown to give a sense of the uncertainty in
the estimates of ages and extinctions.  A green cross indicates the best-fitting model and, for comparison,
an orange cross indicates the values obtained from the nebular emission as described in the previous section.
Overall, there is good agreement between the physical properties derived from the nebular emission
and SED fitting.

\subsection{Integrating Results from the Nebular and Stellar Emission}\label{synth}

By combining results from the nebular and stellar emission,
a self-consistent picture relating the observed properties of
young emerging massive star clusters to model predictions is beginning to develop.
It is noted that caution must be applied when interpreting the physical properties of
observed clusters derived from population synthesis models, since stochastic fluctuations
can bias the results.  However, we do not believe this is significantly affecting the results
presented in this paper since the cluster masses are $\sim 10^4$~M$_\odot$ and we are not
solely relying on broad-band colors \citep{Lancon00,Cervino04}.  In this section,
we combine the results from the nebular (\S\ref{nebprop}) and stellar (\S\ref{sedprop})
emission and explore various trends in the data.
  
\begin{figure}
\begin{center}$
\begin{array}{cc}
\includegraphics[width=3.2in]{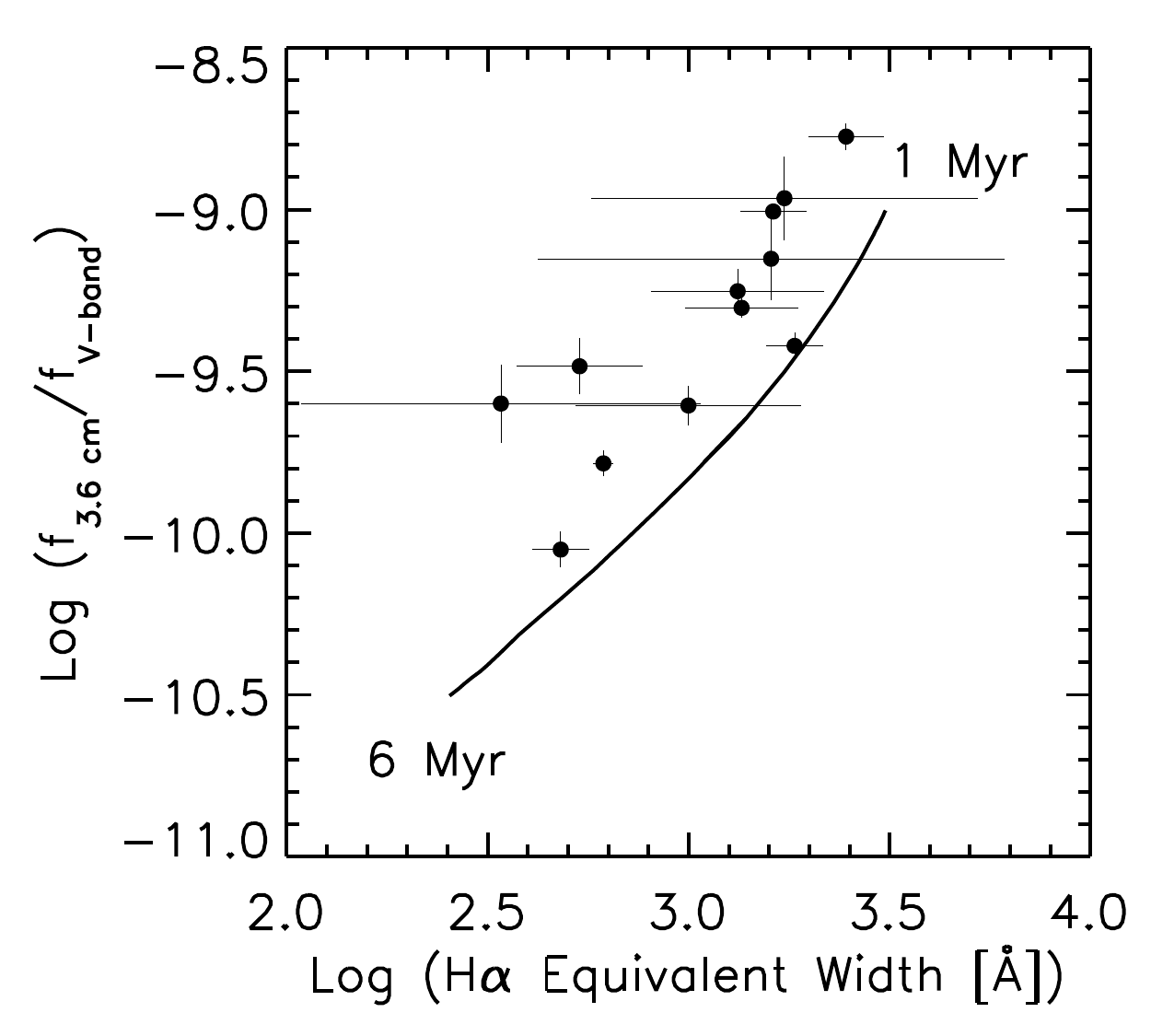} \\
\includegraphics[width=3.2in]{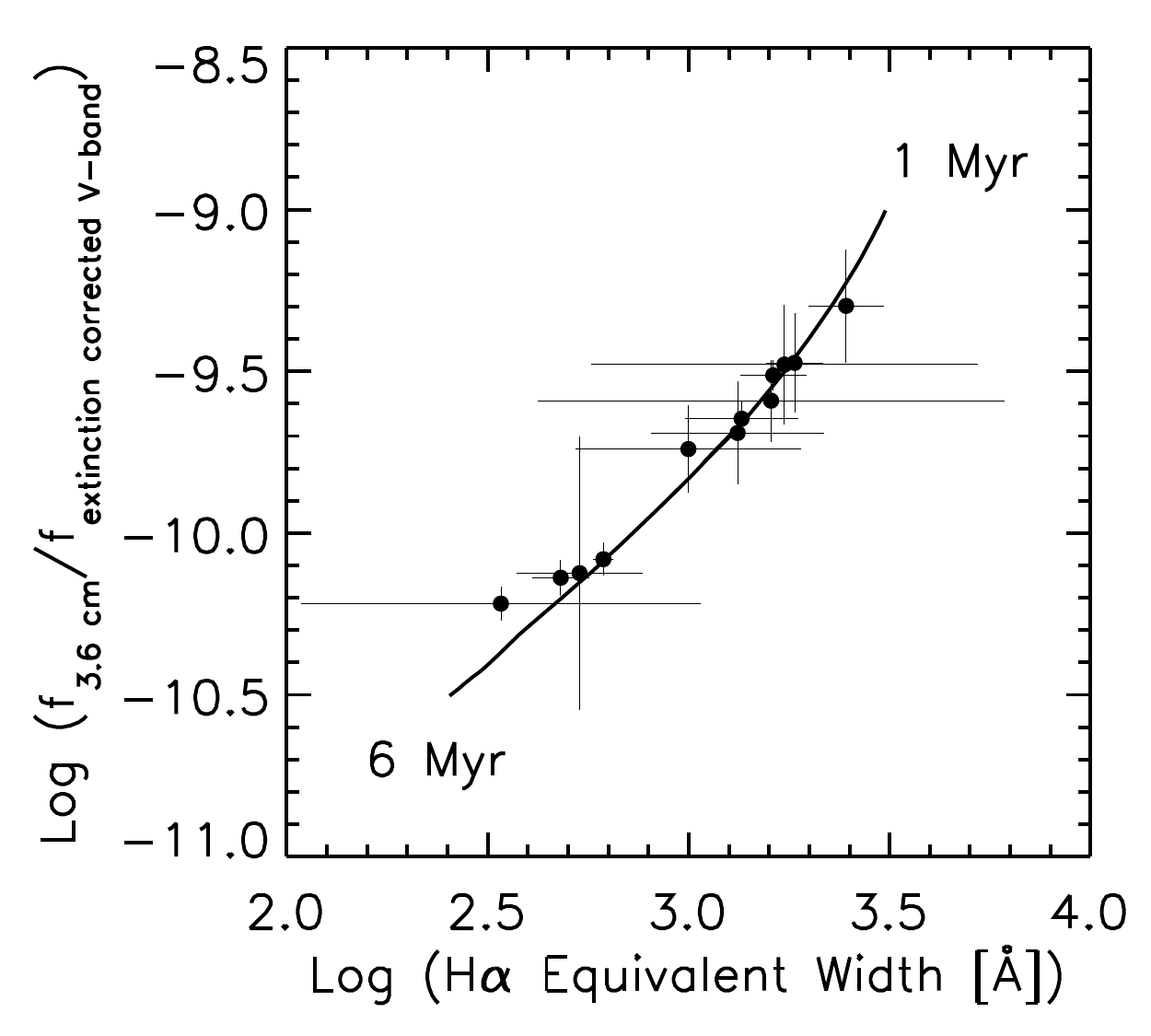}
\end{array}$
\end{center}
\caption{{\it Top:} Ratio of the radio flux densities (3.6 cm) to the
optical ({\it V}-band) flux densities plotted against H$\alpha$ equivalent width
for the thermal sources in NGC~4449.  {\it Bottom:} Same as above except the
{\it V}-band flux densities have been corrected for extinction using the estimates
derived from the nebular gas.  These data match the model evolutionary track extremely well,
illustrating that ionizing flux, and therefore H$\alpha$ equivalent width, do indeed
decrease as a cluster ages (as predicted).  The model flux density
at 3.6~cm was obtained from inserting the predicted ionizing flux from {\scriptsize STARBURST}99
into Equation~\ref{Qlyc}.  These plots also suggest that the extinction
is approximately the same to the stars and the gas.
\label{xvHa}}
\end{figure}

Figure~\ref{xvHa} shows the ratio of radio flux densities (3.6 cm) to the
optical ({\it V}-band) flux densities plotted against H$\alpha$ equivalent width.
The ionizing flux, and therefore H$\alpha$ equivalent width, 
are expected to decrease as a cluster ages.  However, the ionizing flux is also a function
of a cluster's mass whereas H$\alpha$ equivalent width is roughly mass independent.
Normalizing the radio flux densities (which traces the ionizing flux) by the optical
flux densities removes the mass dependence.  It is clear that the mass normalized radio flux densities
and H$\alpha$ equivalent widths do indeed decrease with cluster age.  
A model evolutionary track from 1--6~Myr is also shown in Figure~\ref{xvHa}.  The model flux density
at 3.6~cm was obtained from inserting the predicted ionizing flux from {\scriptsize STARBURST}99 into
Equation~\ref{Qlyc}. The vertical offset between the data and the model evolutionary track can be attributed to
extinction in the {\it V}-band (top of Fig.~\ref{xvHa}).  The data match the model track extremely well after correcting
for the extinctions derived from the nebular emission (bottom of Fig.~\ref{xvHa}).  This result suggests that
the extinction to the stars and the gas is approximately the same, which is significant
since the stellar light and nebular gas may have different geometries (i.e. the nebular emission
is typically more extended than the stellar emission).

\begin{figure}
\begin{center}
\includegraphics[scale=0.65]{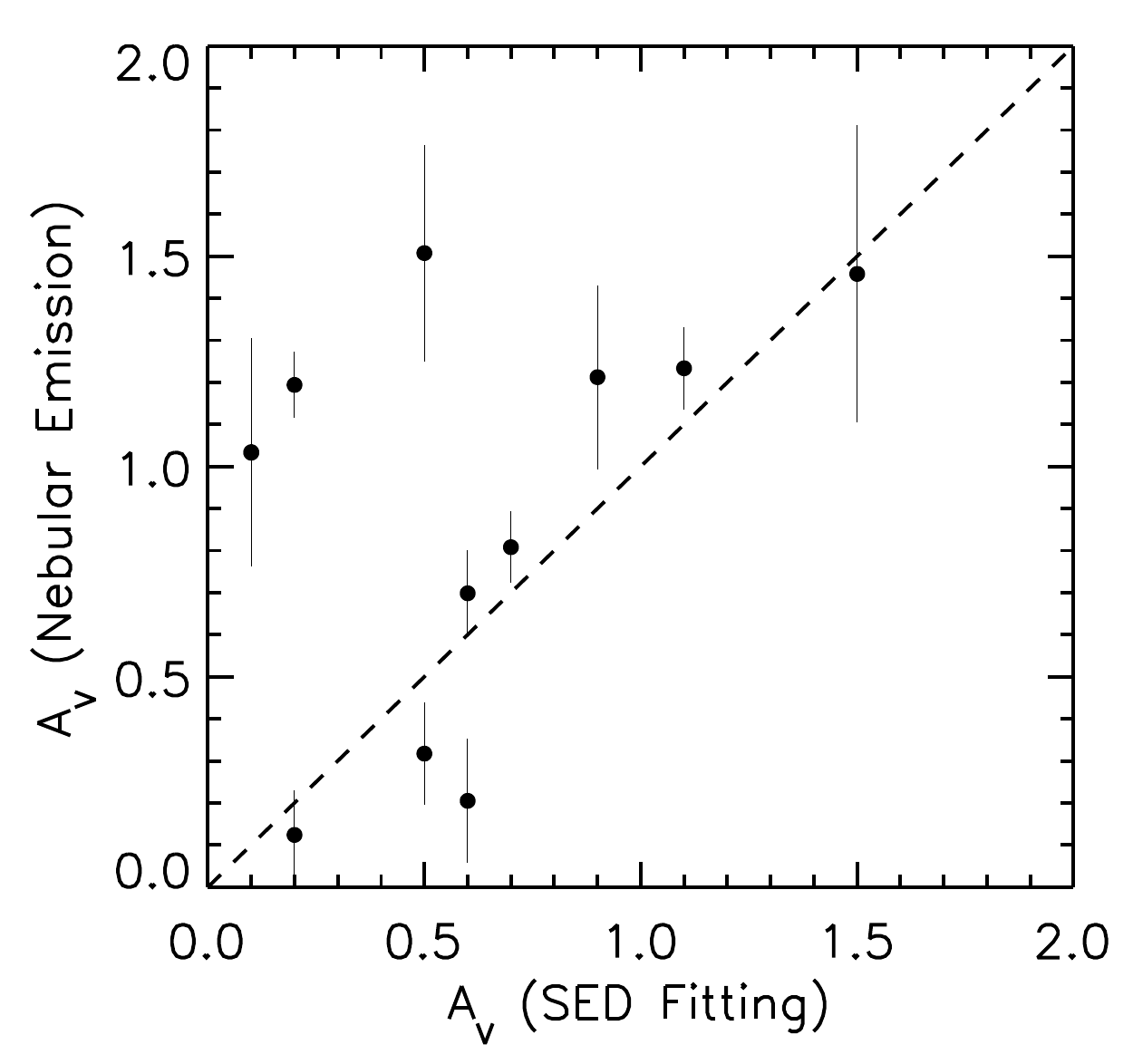}
\caption{Comparison of the extinction estimates of the thermal radio sources in NGC~4449 derived from the
nebular emission and SED fitting.  A line showing the one-to-one correlation is also shown, illustrating
that many sources have similar extinction estimates, yet some sources have a derived nebular extinction that is
$\sim$ 1 magnitude higher than the stellar extinction.  This discrepancy may simply be due to uncertainties in the
estimates from SED fitting.\label{av_thermal}}
\end{center}
\end{figure}

Figure~\ref{av_thermal} also indicates that the extinction to the stars and the gas is roughly
the same in most cases.  The majority of the thermal radio sources have consistent extinction estimates
for the stars and the gas, although approximately one-third of the sources have nebular extinctions
that are $\sim$~1 magnitude higher than their derived stellar extinctions.  Based on the
results from Figure~\ref{xvHa}, these differences may simply be due to uncertainties
in the extinction estimates from SED fitting, which are largely dependent on the ultraviolet
(and lowest quality) data.  The SEDs also suffer a degeneracy between age and extinction.

However, it is worth mentioning here that discrepancies in extinction estimates of extragalactic
\HII\ regions have been recognized since the 1970's.  In particular, early Westerbork data for
\HII\ regions in M33, M101 and M51 \citep{Israel74, Israel75,vanderKruit77} 
showed that comparing radio continuum emission to H$\alpha$ implied about 1.2 mag
excess extinction as compared to the Balmer decrement procedure.  These results combined
with other determinations were summarized by \citet{Israel80}, who discuss
two possible origins for the ``missing dust'': 1) the dust may be internal and mixed with
the ionized gas, 2) over large sizes of many hundreds of parsecs, a non-uniform distribution
of gas and dust will cause the Balmer decrement to be mainly weighted towards regions with
lower extinction, while the radio derived absorption will be determined mainly by the more
absorbed regions.  It is unclear whether the discrepent extinctions derived for 4 of the thermal
radio sources in this work are a result of these types of effects, or simply
due to uncertainties in the SED derived extinctions.

\begin{figure}
\begin{center}
\includegraphics[scale=0.65]{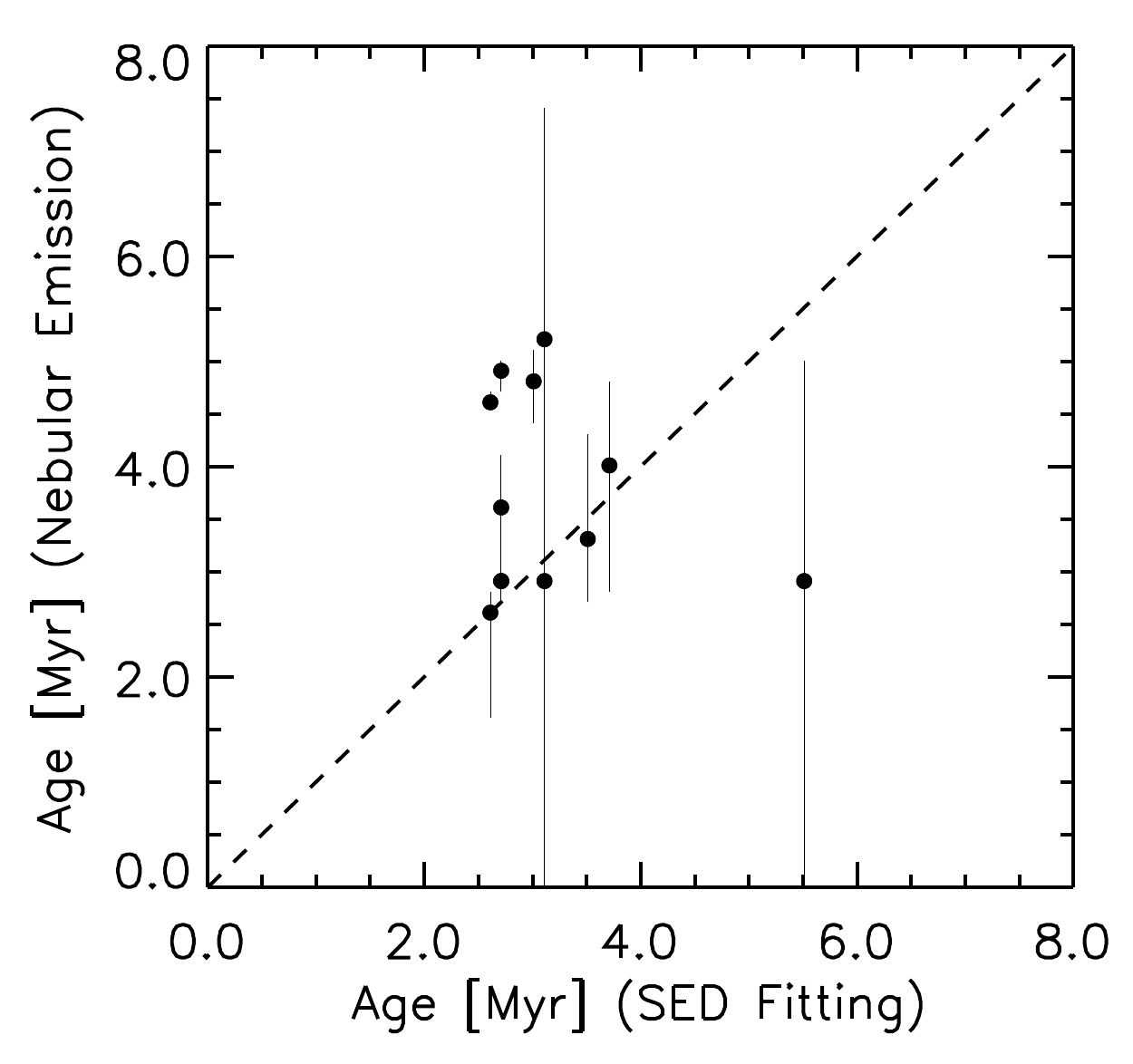}
\caption{Comparison of the age estimates from H$\alpha$ equivalent widths and SED
fitting (and the line showing a one-to-one correlation).  H$\alpha$ equivalent width
is a more sensitive age indicator, since ionizing flux is a strong function of time
once the most massive stars begin to die.  The SEDs, on the other hand, do not change very much at these
young ages which can explain the pile-up at $\sim$~3~Myr.\label{age_thermal}}
\end{center}
\end{figure}

The age estimates from SED fitting also have relatively high uncertainties (see Fig.~\ref{sedthermal}),
and tend to be concentrated around $\sim$~3 Myr.  These results are probably due to an
insensitivity of the model SEDs, which do not change significantly at these young ages.
H$\alpha$ equivalent width, however, is a more sensitive age indicator at young ages
since ionizing flux is a strong function of time once the most massive stars begin to die
(although these age estimates may be systematically high for reasons discussed in \S\ref{ages}).
A comparison of the age estimates from SED fitting
and H$\alpha$ equivalent widths is shown in Figure~\ref{age_thermal}.

\begin{figure}
\begin{center}
\includegraphics[scale=0.6]{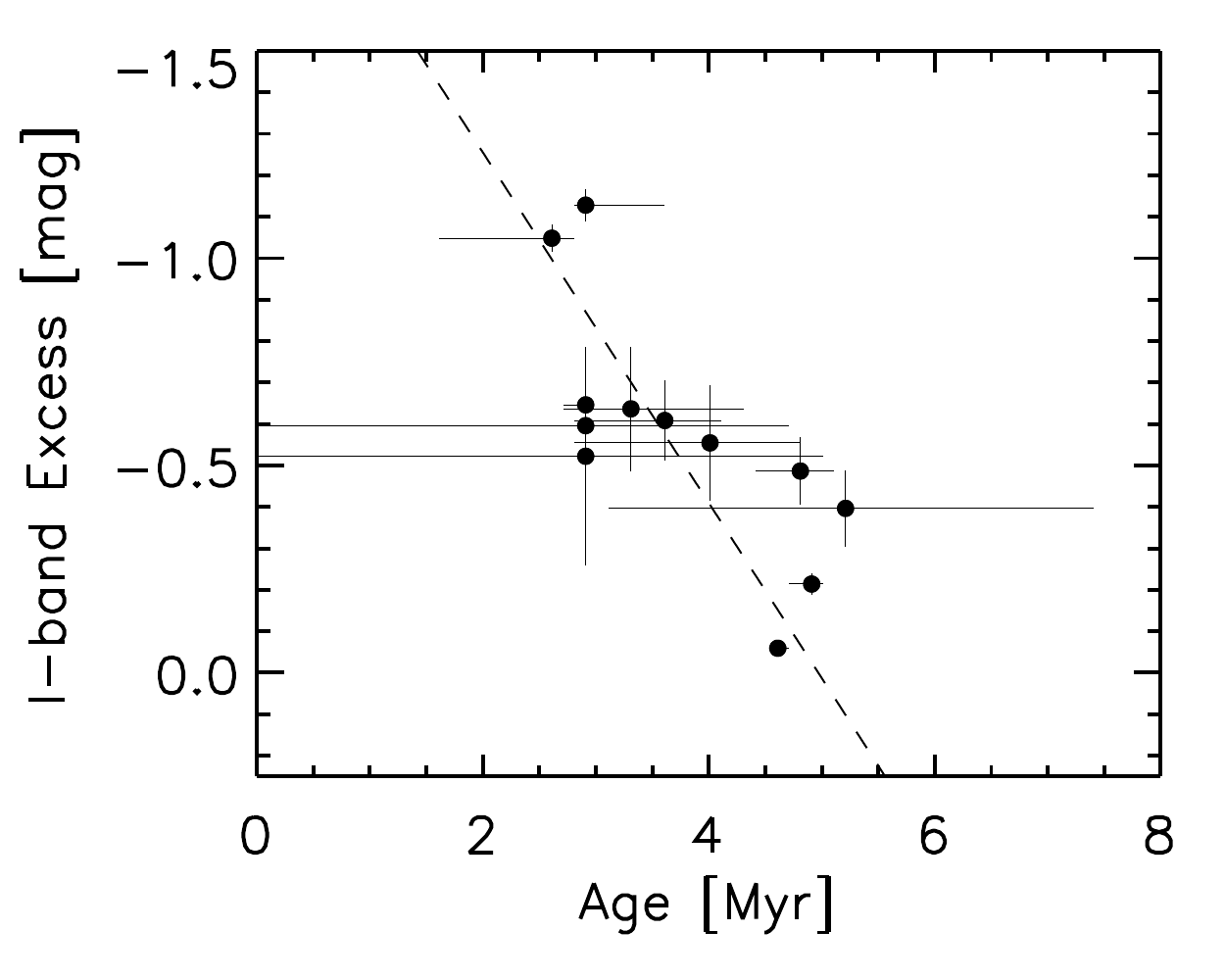}
\caption{Difference between the observed and model {\it I}-band magnitudes versus age,
demonstrating that younger sources have a larger {\it I}-band excess.
The best linear fit to the thermal sources is shown and given by Equation~\ref{Ieqn}.
Based on these data, we do not expect the cause of the observed {\it I}-band excess to affect
clusters older than $\sim$~5~Myr.
\label{Iage_thermal}}
\end{center}
\end{figure}

We find two interesting correlations with H$\alpha$ equivalent width, and hence cluster age. 
First, the {\it I}-band excess (\S\ref{sedprop}) is largest in the youngest clusters and
decreases with age.  We define the {\it I}-band excess in magnitudes as 
$\Delta I = -2.5 \times {\rm log}(f^{(I)}_{\rm observed}/f^{(I)}_{\rm model})$,
where $f^{(I)}_{\rm observed}$ is the observed flux density in the F814W filter and $f^{(I)}_{\rm model}$
is the flux density of the best fitting model SED.  Figure~\ref{Iage_thermal}
shows a plot of $\Delta I$ versus age for the thermal radio sources and the best linear
fit is given by

\begin{equation}
\left({\Delta I \over {\rm mag}}\right) =  0.4 \left({t \over {\rm Myr}}\right) - 2.1.
\label{Ieqn}
\end{equation}

\noindent
Setting $\Delta I$ equal to zero implies that whatever is causing the {\it I}-band excess
in these young clusters is unlikely to affect clusters older than $\sim$~5~Myr.
We will discuss possible origins of the {\it I}-band excess in \S\ref{irexcess}. 

\begin{figure}
\begin{center}$
\begin{array}{ccc}
\includegraphics[width=3.3in]{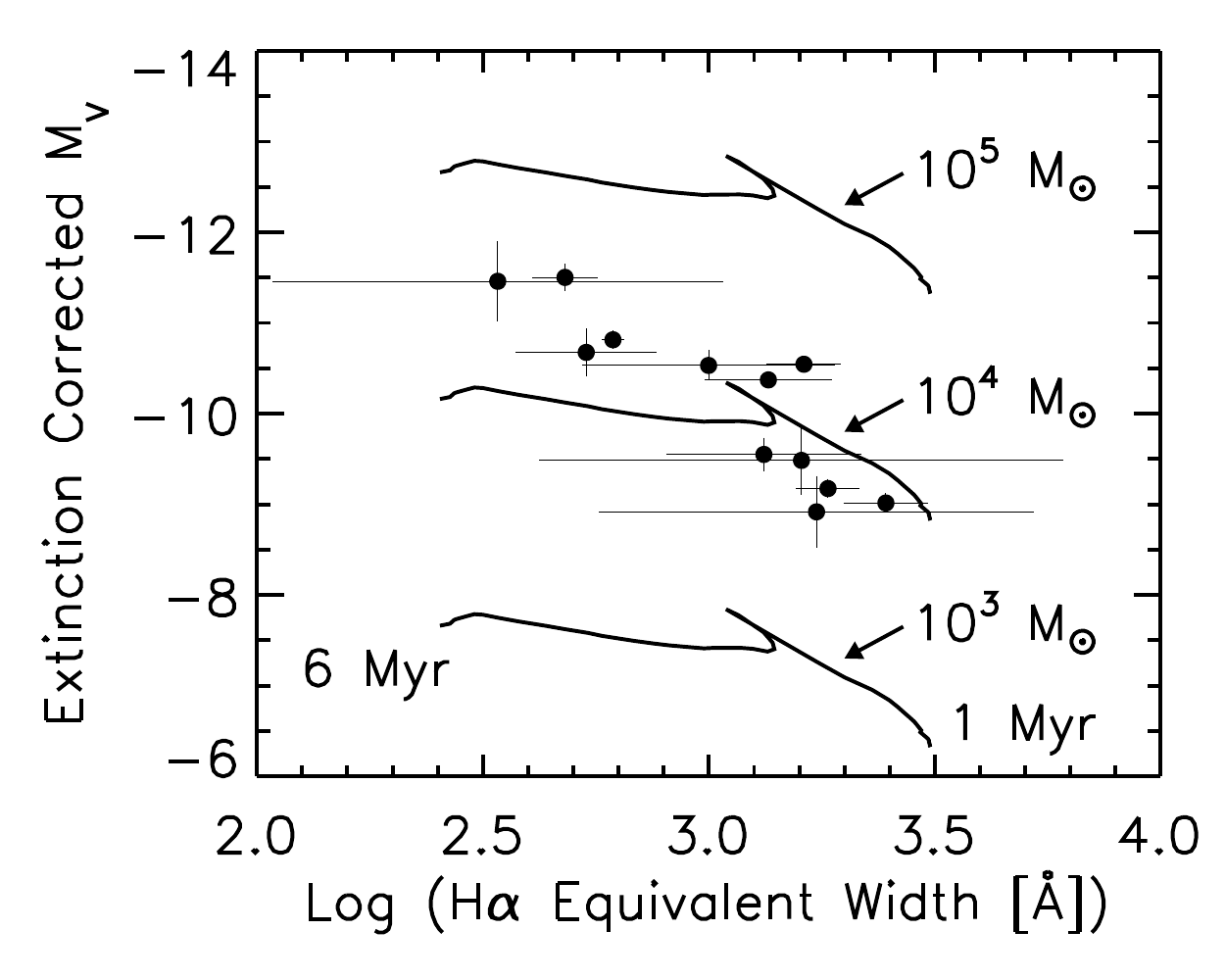} \\
\includegraphics[width=3.3in]{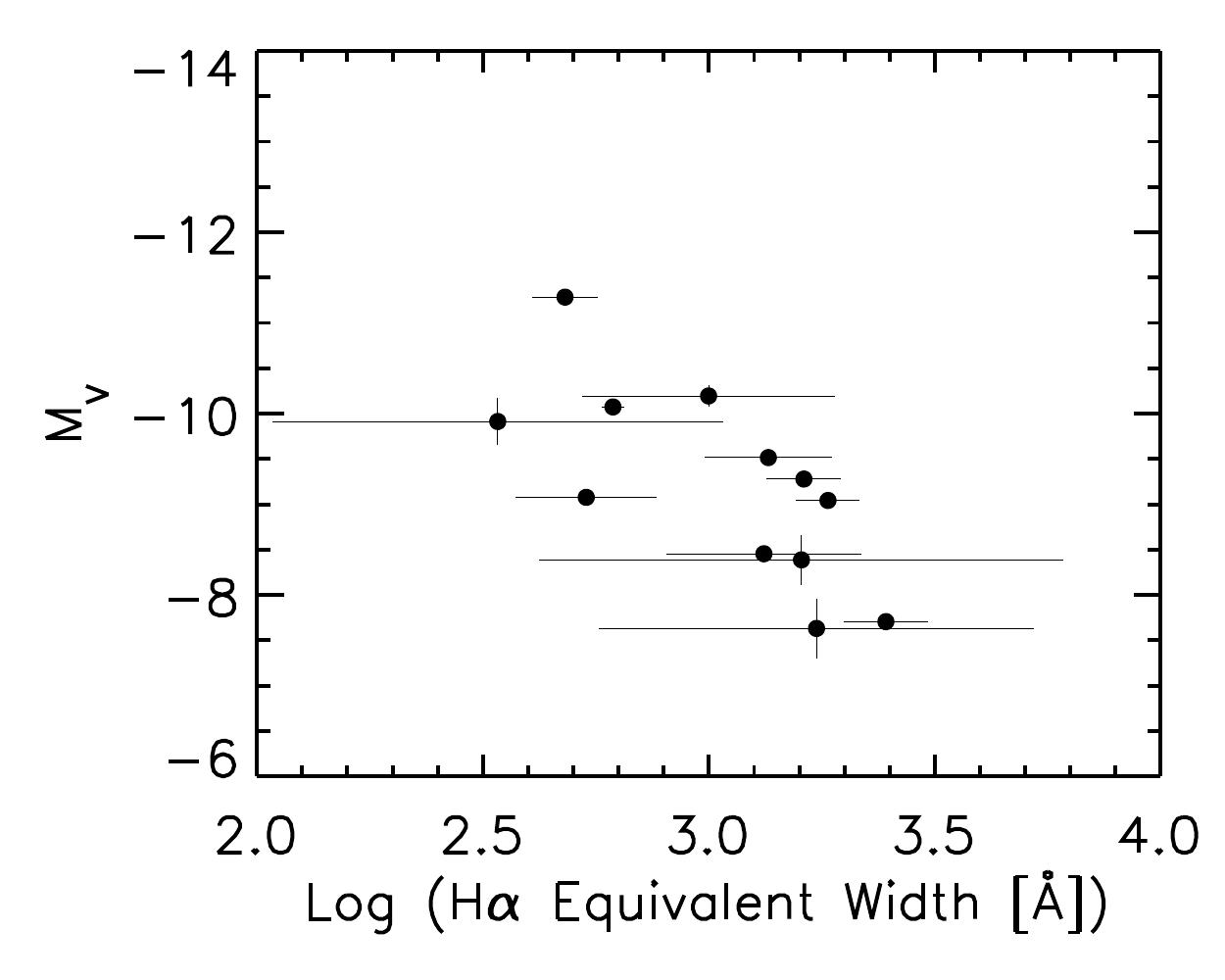} \\
\includegraphics[width=3.3in]{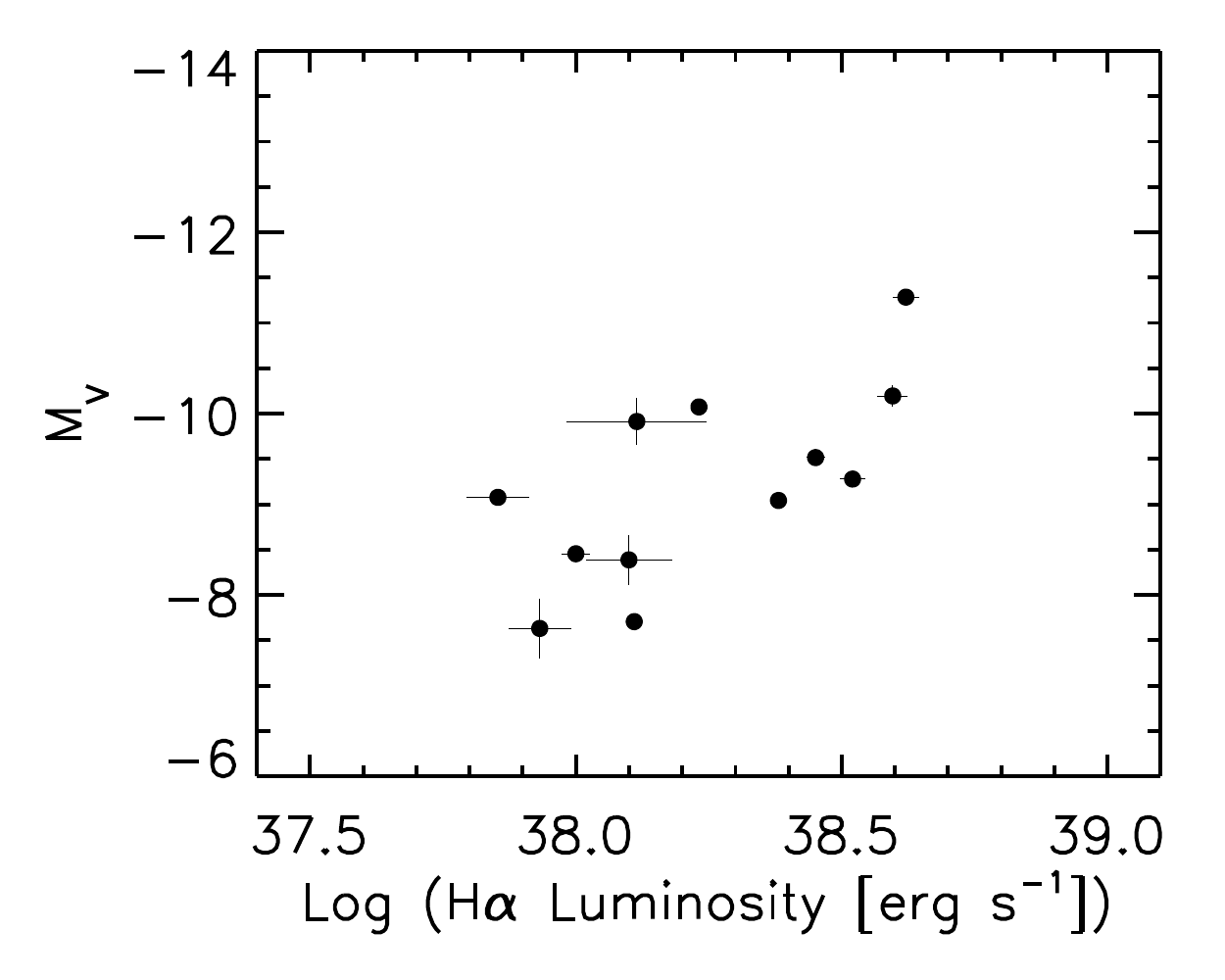} \\
\end{array}$
\end{center}
\caption{{\it Top:} Extinction corrected absolute $V$ magnitude versus H$\alpha$ equivalent width
of the thermal radio sources.  The extinction estimates come from comparing the
radio and H$\alpha$ emission from the ionized gas.  Model evolutionary tracks for a range of masses
are also shown.  This plot suggests there is a correlation between cluster mass and age, in
excess of the model predictions.  See \S\ref{mass_age} for a discussion. 
{\it Middle:} The same overall trend is observed without correcting M$_V$ for extinction.
{\it Bottom:} M$_V$ and H$\alpha$ luminosity are positively correlated, as expected.
\label{deredMvHaEW}}
\end{figure}

We also observe a puzzling trend that seems to indicate cluster mass is positively correlated with age
for ages $\lesssim 5$~Myr.
Figure~\ref{deredMvHaEW} shows a plot of the extinction corrected M$_V$ versus H$\alpha$ equivalent width.
The model evolutionary tracks for various cluster masses are also shown.
A strong anti-correlation between M$_V$ and H$\alpha$ equivalent width is evident,
implying more massive clusters tend to be older than less massive clusters.  The same
overall trend is observed without correcting M$_V$ for extinction, while a positive correlation
is found between M$_V$ and H$\alpha$ total flux (as expected).  An observational bias could possibly explain
why optically faint clusters with low H$\alpha$ equivalent widths are not detected, since these
clusters may be less likely to show up in the radio and make it into our sample.  However, it is
curious that we do not detect clusters that are relatively bright in $V$ with high H$\alpha$ equivalent
widths.  It is not clear what is causing this trend in the data, but
we explore various possibilities in \S\ref{mass_age}.

\begin{figure}
\begin{center}
\includegraphics[scale=0.65]{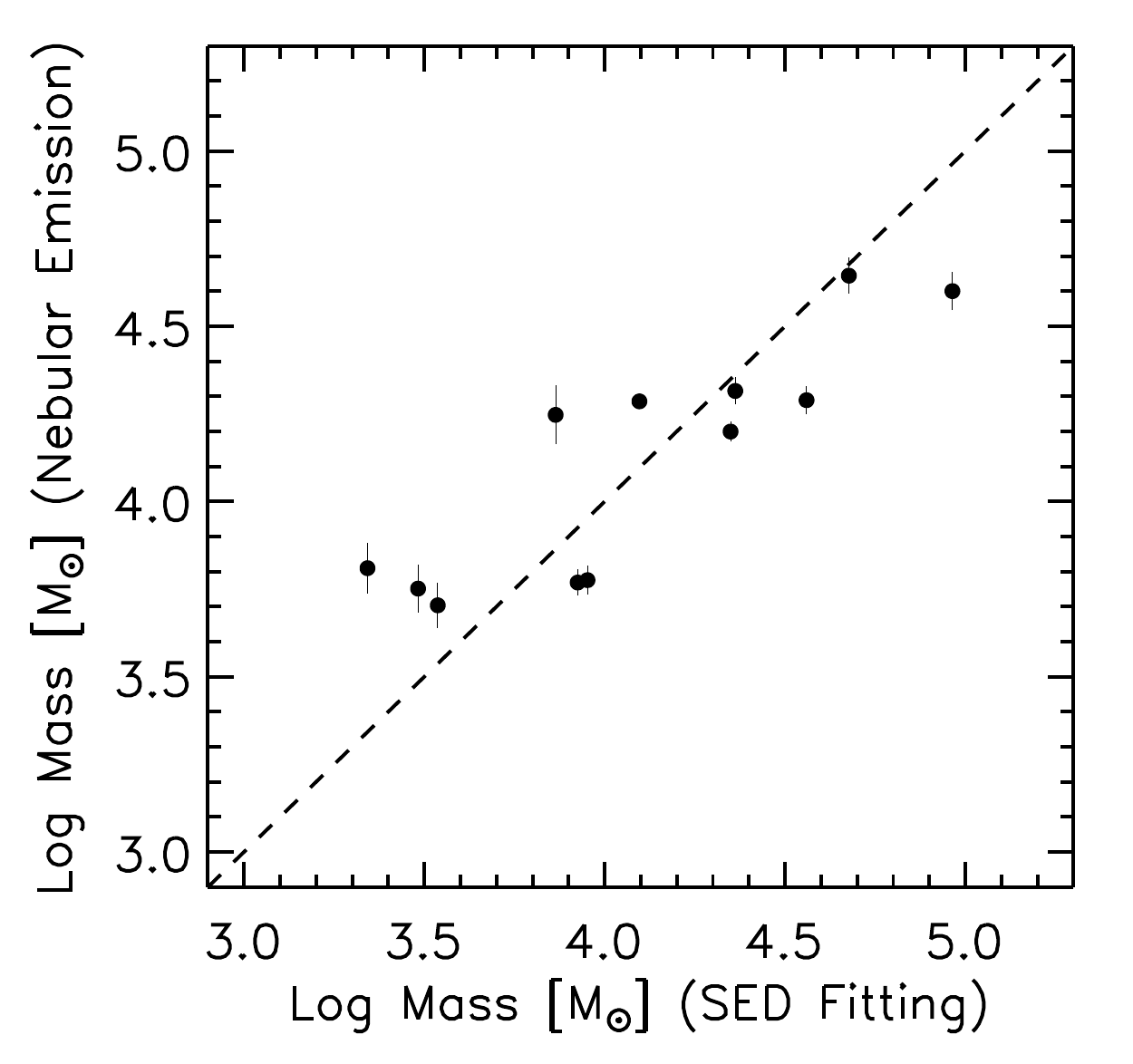}
\caption{Comparison of the stellar masses of the thermal radio sources in NGC~4449 derived from nebular
emission and SED fitting.  The line showing a one-to-one correlation is also shown.  A slight trend
is apparent that suggests mass estimates from the nebular emission tend to be higher than those
from SED fitting for the lower mass clusters, while the mass estimates from SED fitting tend to be
higher than those derived from the nebular emission for the higher mass clusters.
\label{mass_thermal}}
\end{center}
\end{figure}

Overall, the mass estimates derived from the nebular emission and SED fitting are consistent within
a factor of $\sim 3$.  However, a slight trend is evident in Figure~\ref{mass_thermal} suggesting that
mass estimates from the nebular emission tend to be higher than those from SED fitting for the lower mass
clusters, while the mass estimates from SED fitting tend to be higher than those derived from the nebular
emission for the higher mass clusters.  We will return to this in \S\ref{mass_age}

\section{Discussion}

\subsection{Possible Origins of the {\it I}-band Excess}\label{irexcess}

We have demonstrated that the {\it I}-band fluxes of the radio selected clusters in
NGC~4449 exhibit an excess compared to model spectral energy distributions (Fig.~\ref{sedthermal}).
In the $B-V$ versus $V-I$ color-color diagram, the reddest clusters in $V-I$ are simultaneously
the bluest in $B-V$ (Fig.~\ref{BVvsVI}).  The {\it I}-band excess is also positively correlated
with H$\alpha$ equivalent width, implying the excess decreases with age (Fig.~\ref{Iage_thermal}).
Based on these observations, we do not expect the origin of this {\it I}-band excess to affect clusters
older than $\sim$~5 Myr (Eqn.~\ref{Ieqn}).  

Possible origins of the {\it I}-band excess are discussed below, but first we note that
other studies have also revealed the presence of an {\it I}-band
excess in a small number of young massive clusters.
\citet{Whitmore95} discovered about a dozen extremely red objects in the SE quadrant
of the merging Antennae galaxies.  This region also contains the strongest radio continuum sources
in the system as well as strong mid-IR emission and a few of these sources are probably associated
with the red objects.  \citet{Whitmore95} interpreted these red objects as extremely young,
embedded clusters that have not yet cleared out their natal dust cocoons.

In their study of the circumnuclear starburst galaxies NGC~5248 and NGC~1512,
\citet{Maoz01} found a number of young massive clusters with excess emission in the {\it H}-band,
and one of these sources had a detectable {\it I}-band excess.  \citet{Maoz01} were able to fit the SED of this
source by adding a 2000 K blackbody to their model and determined that {\it if} the emission were due to
dust heated by stars, it must be circumstellar.  

Possible origins of the observed {\it I}-band excess in the young radio-detected clusters of NGC~4449
are considered here.

\vspace{.35cm}

\noindent
i) {\it Red Supergiants}

One possible origin for an {\it I}-band excess is the presence of red supergiants.
These evolved stars are known to dominate the near-infrared light of massive clusters
older than $\sim$~7~Myr \citep[e.g.,][]{McCrady03}, but we cannot definitively rule out
their presence in the younger clusters of NGC~4449 because of stochastic variations.

In addition, since we use the 3.6 cm radio image as our reference when doing photometry,
the sizes of our apertures are larger than the linear resolution of the {\it HST} images and
individual compact star clusters.  Some of our apertures contain multiple discrete sources, of which
some are very red and could be individual supergiants. 

On the other hand, there are radio sources with single dominant optical counterparts.  For example,
Source 30 is a compact optical cluster with a linear size $\lesssim 3$~pc (FWHM), and it
exhibits the largest {\it I}-band excess (and youngest age).  It would be quite
unexpected for this young ($<$ 3 Myr) cluster to contain red supergiants.  In addition, the
{\it I}-band excess {\it decreases} with cluster age (Fig.~\ref{Iage_thermal}) while the presence of supergiants
should increase with time for clusters less than 10 Myr old.  We therefore consider it highly unlikely that
red supergiants are the primary cause of the {\it I}-band excess.

\vspace{.35cm}

\noindent
ii) {\it Emission Lines}

Emission lines are another possible source of the observed {\it I}-band excess.  The Paschen series
(9 and higher) and the [SIII] $\lambda$9069, $\lambda$9532 emission lines could contribute to the flux
in the F814W filter and these lines have all been observed in the spectra of \HII\ galaxies and starbursts
\citep[e.g.][]{Kehrig06,Marquart07}.

To see if emission lines could plausibly be the dominant source of the {\it I}-band excess, we   
calculate the extinction corrected flux ratio, relative to H$\alpha$, needed
to account for the excess flux in the {\it I}-band and find it to be $\sim$~0.55 on average with a range between
$\sim$~0.1--1.  In other words, the requisite contribution from emission lines in the F814W filter would need
to be $\sim$~10--100\% of the H$\alpha$ flux for a given source.  The higher level Paschen
lines are not expected to contribute nearly this much flux (e.g. Pa9/H$\alpha < 1\%$).
The ratio of the [SIII] lines to H$\alpha$ has been observed to be as high as     
20--30\% \citep{Kehrig06} and may play some role here,
but the integrated system throughput for the F814W filter drops by 50--75\% from its
maximum at the wavelengths of these lines.  For all of these reasons, we believe that emission lines are
probably not the dominant source of the {\it I}-band excess.

\vspace{.35cm}

\noindent
iii) {\it Thermal Emission from Hot Dust}

It is possible that thermal emission from hot dust near the sublimation temperature ($\sim$~1700 K)
could contribute to an excess in the {\it I}-band.  An excess at $\sim~2~\mu$m is observed in many
star forming regions and usually attributed to hot dust.  
However, the youngest radio detected clusters in NGC~4449 have such large
{\it I}-band excesses that it would be truly remarkable if thermal emission from heated dust were the
dominant cause since either there must enough emitting dust near the sublimation temperature to
dominate the stellar light at $\sim~0.8~\mu$m or the dust must survive even higher temperatures.

\vspace{.35cm}

\noindent
iv) {\it Optically Veiled Photospheric Continuum from Embedded Stars}

A sub-population of deeply embedded stars that have not yet fully emerged from their protostellar
clouds could also potentially cause an {\it I}-band excess.  These stars would be invisible at
optical wavelengths because of high extinction, but their photospheric emission could
contribute to the integrated cluster light in the near-IR and even in the {\it I}-band.
However, the observed {\it I}-band excess is so large in some sources with no obvious excess in the
{\it V}-band, that the embedded population would be required to simultaneously have moderate to
large extinctions ($A_{\rm V} \sim 5$) and uncomfortably high fractions ($\sim$~90\%) of the
total cluster mass.

\vspace{.35cm}

\noindent
v) {\it ``Extended Red Emission''}

Extended Red Emission (ERE) is a very promising explanation for the {\it I}-band excess
(see \citet{Witt04} for a review).  ERE is typically observed as a
broad emission feature ($\sim 1000$~\AA) which peaks in the red to near-IR spectral range.
ERE has been discovered in number of astrophysical environments where both interstellar dust
and ultraviolet photons are present, including the Orion Nebula \citep{Perrin92} and the
30 Doradus Nebula \citep{Darbon98}.

ERE is caused by photoluminescense, a 3-step process in which 1) an interstellar
particle absorbs an ultraviolet photon and makes an electronic transition to an excited state, 2) a series of
rotational/vibrational transitions relaxes the system to an intermediate state, and 3) an
optical photon is emitted through an electronic transition back to the ground state.
The carrier of ERE is unknown.

Studies of ERE have revealed that, in any given environment, the maximum intensity is well
correlated with the local density of the radiation field \citep{Gordon98}, and in particular with
the density of the {\it ultraviolet} radiation field \citep{Smith02}.  The ERE peak wavelength is also
closely correlated with the density of far-UV photons, suggesting that the radiation field is not
only driving the ERE, but also modifying the carrier through size-dependent photo-fragmentation,
preferentially affecting the smallest carrier particles first \citep{Smith02}.  Furthermore,
nebulae in which ERE is not detected \citep{Witt90} tend to be diffuse (unclumped) and have had exposure
to UV radiation for an extended amount of time.  In environments with clumpy dust, ERE is strongest
on the faces of high-density clumps and undetectable in low-density interclump regions.

These findings are consistent with ERE causing the observed {\it I}-band excess in the young
massive star clusters in NGC~4449 (recall that the excess decreases with cluster age and is expected
to disappear after $\sim 5$~Myr).  The youngest clusters will have the most intense UV radiation fields
and they are also the most likely to have a clumpy dust structure, both of which would increase the
intensity of the ERE and hence the {\it I}-band excess.  Over time, the clumpy dust structure may be
dispersed and/or the radiation field may modify/destroy the ERE carrier, both causing the {\it I}-band
excess to decrease and then disappear all together.  

ERE offers a favorable hypothesis for the {\it I}-band excess, although spectroscopic
follow-up is required for confirmation.  It is also important to note that if ERE is the
cause of the {\it I}-band excess, the {\it H}-band excess likely has a different origin.
Contributions from hot dust, embedded stars and red supergiants are all possible.  
We are currently undertaking follow-up spectroscopic and photometric observations of NGC~4449
to further investigate the {\it I}- and {\it H}-band excesses.

\subsection{Interpretation of the Mass-Age Correlation for Ages $\lesssim$ 5 Myr}\label{mass_age}

Figure~\ref{deredMvHaEW} illustrates that while M$_V$ and H$\alpha$ flux are positively
correlated (as expected), M$_V$ and H$\alpha$ equivalent width are anti-correlated.
In other words, fainter clusters have relatively more H$\alpha$ emission with respect to the
stellar continuum than brighter clusters. 
This can be interpreted as a mass-age correlation, since M$_V$ generally traces cluster mass (modulo extinction)
and H$\alpha$ equivalent width traces age.
The two most trivial, and unsatisfying, interpretations of this trend are 1) the low-mass clusters
formed more recently than the high-mass clusters, and 2) smaller clusters coalesce and make larger clusters
over time.  However, we note that there is no observed correlation between cluster size (measured by aperture size)
and H$\alpha$ equivalent width.

A third interpretation of the apparent mass-age correlation for clusters in this work (with ages
$\lesssim 5$~Myr), although purely speculative, is
the following.  Is is possible that the observed correlation is actually the imprint of an evolutionary sequence,
where the youngest clusters merely appear less massive than their older counterparts.
First, relative to more evolved clusters, younger clusters are likely to contain a higher fraction
of dust enshrouded, optically veiled stars.  These heavily embedded stars would
not contribute to the integrated {\it V}-band flux, yet the young optically visible stars within
the clusters would still produce large H$\alpha$ equivalent widths revealing their young ages.
Therefore, younger clusters may tend to appear less massive.  In addition, if the younger
clusters have a majority of their mass in the optically veiled sub-population of stars, it would not
be surprising that the masses inferred from the nebular emission (i.e. radio and H$\alpha$ equivalent
width, which are mostly unaffected by extinction) are higher than those inferred from the stellar
emission (which only samples the optically visible ``emerged'' stars) as is apparent in
Figure~\ref{mass_thermal}.

A second factor that could contribute to the apparent mass-age correlation is
stellar winds.  More massive clusters will have disproportionately
strong stellar winds, which could preferentially clear away the gas.  A
lower gas fraction will result in a lower equivalent width, and therefore
a larger inferred age.  As a result, more massive clusters would tend to
appear older by virtue of their dearth of ionized gas.  A deficiency of ionized gas
could also explain why the more massive clusters tend to have higher mass estimates from
SED fitting than those derived from the nebular emission (Fig.~\ref{mass_thermal}).

\subsection{Evidence for Triggered Star Formation in NGC~4449}\label{trigger}

NGC~4449 is thought to have undergone some type of interaction in its past
\citep{Hunter98,Theis01}, which has likely influenced the global star formation
history of the galaxy over the last $\sim 1$~Gyr \citep[e.g.][]{Annibali07}.  
Local processes such as stellar winds and supernovae (perhaps the result of a previous burst), however,
seem to be playing a significant role in triggering the {\it current} bursts of
star formation, resulting in the formation of massive clusters detected in the radio.

\begin{figure}
\begin{center}
\includegraphics[scale=0.4]{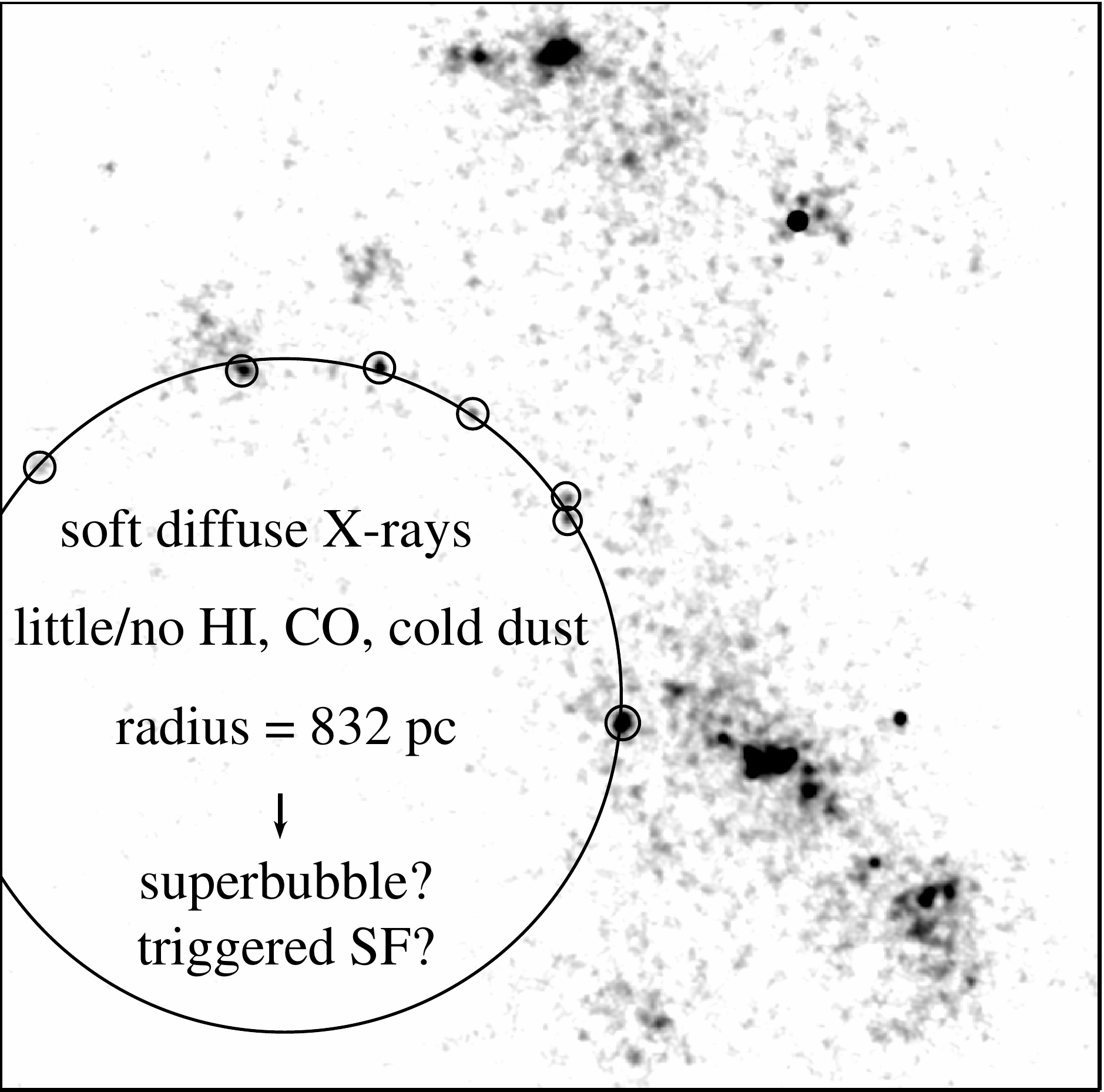}
\caption{VLA 3.6 cm radio image of NGC~4449.  A 1.6 kpc superbubble may have triggered
the formation of the 7 radio sources at the circle's perimeter.  See \S\ref{trigger} for more
discussion.\label{superbubble}}
\end{center}
\end{figure}

Figure~\ref{superbubble} illustrates that the formation of a number of radio sources appears
to have been triggered by an expanding superbubble.
Taken alone, the fact that seven sources (23, 24, 25, 28, 30, 37, \& 39) lie along the
same circular arc (radius = 44\arcsec = 832 kpc) could be viewed
as a coincidence, albeit an unlikely one.  However, previous observations of gas
and dust in NGC~4449 suggest otherwise.  \HI, CO and cold dust appear to be depressed or
almost absent in this region of the galaxy \citep[e.g. Fig.~1 in][]{Bottner03}
while soft diffuse X-ray emission fills it in
(Fig.~5 in \citet{dellaCeca97} and Fig.~10 in \citet{Summers03}).
In addition, kinematic studies of NGC~4449 are consistent
with the stellar component of the galaxy observed face-on \citep[][2005]{Hunter02},
which would account for the sources lying on a circular arc in projection.
All of this evidence supports the view of an expanding superbubble
triggering several sites of star formation in NGC~4449.

Further evidence of triggered star formation is apparent within NGC~4449.
An H$\alpha$ shell with radius $\sim 250$ pc is observed in the SW region of the galaxy (see
Figs.~\ref{multiwave_ims} \& \ref{closeups}) surrounding numerous optically visible
stars and clusters, whose stellar winds and supernovae likely produced the shell,
while six radio sources (1, 2, 3, 4, 5, \& 7) are found along the periphery of this shell.
It is possible that a collision between this shell and the larger superbubble is
responsible for compressing the gas in the southern region of the galaxy and triggering
the formation of Sources 18, 20, and 21.

Another group of radio sources (31--36) is found along a smaller H$\alpha$ loop (radius $\sim 40$ pc) 
in the NE region of the galaxy, again containing optically visible stellar
clusters that could have triggered the current local burst of star formation.

\subsection{Radio Sources without Optical or Infrared Counterparts}\label{pureradio}

A close inspection of Figures~\ref{multiwave_ims} \& \ref{closeups} reveals that at least
a few radio sources in NGC~4449 do not have optical counterparts (e.g. Sources 20, 24, \& 28 ).
Sources 24 \& 28 also do not appear to have any detectable 24~$\mu$m emission associated with them.
It is unclear whether this is the case for the other radio sources without optical counterparts
due to confusion in the 24~$\mu$m image.
In addition, the ``pure'' radio sources have only been detected at 3.6 cm and we
cannot determine whether they are thermal.  We do believe these sources are in the galaxy and not background
objects because their locations are associated with star forming regions.  
With the current data, however, we can only speculate about the nature of these sources.

If the sources are thermal, the lack of H$\alpha$ could be explained by heavy extinction from dust.
For Sources 24 \& 28, the extinctions would have to be at least $\sim$~7 visual magnitudes
(\S\ref{extinction}).  Again, if we assume these sources are thermal, they are probably
younger than the radio detected clusters with optical counterparts ($\lesssim~2.5$~Myr).  They would also have
relatively low ionizing fluxes (the equivalent of $\sim$~6--15 O7.5 V stars) and masses.  These young ages and low
masses would be consistent with the observed correlation between derived masses and ages for the thermal radio
sources {\it with} optical counterparts.  However, if the visibly obscured sources are heavily embedded clusters,
we would expect them to produce thermal emission from heated dust and be observable in
the mid-IR which does not seem to be the case (at least for Sources 24 \& 28).  It may be more likely that
these ``pure'' radio sources are supernova remnants, since non-thermal emission is anti-correlated
with bright mid-IR emission \citep[e.g.][]{Brogan06}.
  
\section{Summary and Conclusions}

We have presented a multi-wavelength study of emerging massive star clusters in the irregular starburst
galaxy NGC~4449.  We combine sensitive, high resolution VLA radio observations with ultraviolet,
optical and infrared data from the {\it Hubble} and {\it Spitzer Space Telescope} archives to
obtain a comprehensive view of star formation occurring within the galaxy and to determine
the physical properties of the radio-detected star clusters.
We summarize the results of our study below:

\begin{enumerate}

\item We identify 13 thermal radio sources in NGC~4449.  These sources have ages
$\lesssim 5$~Myr (based on H$\alpha$ equivalent widths), ionizing fluxes
between $\sim$~20--70~$\times 10^{49}$~s$^{-1}$ (the equivalent of $\sim$~20--70 O7.5 V stars),
and stellar masses between $\sim$~0.5--5~$\times 10^4$~M$_\odot$.

\item Of the 13 thermal radio sources, 12 have {\it measured global} extinctions $\lesssim 1.5$ magnitudes
at 5500~\AA~and the extinction appears to be approximately the same to the stars and the gas.  The most
optically obscured thermal radio source has an $A_{\rm V} \approx 4.3$.  

\item Model SEDs modified by a 30 Doradus extinction curve fit the clusters' broad-band ultraviolet and
optical flux densities ($UV, U, B, V$) significantly better than the standard Galactic extinction
curve and the starburst obscuration curve of \citet{Calzetti00}.

\item The clusters exhibit an {\it I}-band excess (with respect to model SEDs) that is
anti-correlated with age.  In the most extreme cases, the {\it I}-band magnitudes are more
than 1 magnitude brighter the best-fitting model SEDs.  We do not expect clusters with ages
$\gtrsim 5$~Myr to have this excess.  A photoluminescent process, known as Extended Red Emission,
provides a favorable hypothesis for the cause of the observed {\it I}-band excess.

\item The 7 nuclear clusters with NICMOS data exhibit a large ($\sim 1.6$ mag) {\it H}-band excess.
We speculate that some combination of red supergiants, thermal emission from hot dust, and optically veiled
photospheric emission are the most likely causes.

\item An apparent mass-age correlation (for ages $\lesssim 5$~Myr) is observed for the thermal
radio sources in NGC~4449.

\item The mid-infrared (24~$\mu$m) morphology of NGC~4449 is almost identical to that of
the radio, indicating that nearly all of the warm dust in the galaxy is associated with
the most recent episodes of star formation in the galaxy.

\item Local processes such as supernovae and stellar winds are likely playing an important role
in triggering the {\it current} bursts of star formation within NGC~4449.

\end{enumerate}

Synthesizing high quality data sets across the electromagnetic spectrum has provided
us with a unique and detailed view of the current star formation occurring within NGC~4449.
Our findings would not have been possible without the {\it combination} of radio, optical, and infrared
observations.  The results presented here are the first of a larger program to study
massive star clusters as they emerge from their birth material and
transition from being visible in the radio to optical wavelength regimes.
Ultimately, understanding the formation and early evolution of massive star
clusters will shed light on a fundamental mode of star formation
throughout the universe and provide insight into the origin
of globular clusters.

\acknowledgments

We thank the anonymous referee for numerous insightful comments and suggestions.
A.E.R appreciates useful discussions with Bob O'Connell, Remy Indebetouw,
Crystal Brogan, Rupali Chandar, Brad Whitmore, Nate Bastian, Mike Dopita, Daniella Calzetti,
Zhi-Yun Li, Ricardo Schiavon, Craig Sarazin, John Hibbard, Jarron Leisenring and David Nidever.  
A.E.R. is thankful for support from the Virginia Space Grant Consortium and the
University of Virginia through a Governor's Fellowship.
K.E.J. gratefully acknowledges support for this paper provided by NSF
through CAREER award 0548103 and the David and Lucile Packard Foundation
through a Packard Fellowship.  This work is based in part on archival
data obtained with the Spitzer Space Telescope, which is operated by
the Jet Propulsion Laboratory, California Institute of Technology
under a contract with NASA. Support for this work was provided by an
award issued by JPL/Caltech.  Support for program \#AR09934 was
provided by NASA through a grant from the Space Telescope Science
Institute, which is operated by the Association of Universities for
Research in Astronomy, Inc., under NASA contract NAS 5-26555.

{\it Facilities:} \facility{VLA}, \facility{HST}, \facility{SPITZER}.

\appendix

\section{SURPHOT: A Multi-wavelength Photometry Code}

We have developed a new multi-wavelength photometry code (SURPHOT) to measure
flux densities within identical {\it irregular} apertures in multiple images.  SURPHOT
is written in the Interactive Data Language (IDL) and the basic algorithm is as follows:
The user supplies a reference image and selects an aperture type, including ``contour'' or
``free-form'' (in addition to a circle or an ellipse).  If a contour aperture is chosen, SURPHOT will
detect a specified contour level to define a region of interest.  Alternatively, the
user can choose to draw a free-form aperture by clicking around a region of interest.  Once an aperture is
defined in the reference image, a background annulus is assigned by expanding the aperture out
radially by two specified factors, matching the overall shape of the aperture.  The
pixel coordinates of vertices in the aperture and background annulus are transformed into
right ascension and declination using astrometry keywords in the FITS header of the reference
image.  The RA and DEC values can then be transformed into pixel coordinates of any other image 
using its own FITS header astrometry keywords.  In this way, SURPHOT uses the same aperture
and background annulus to compute flux densities in any number of user supplied images, provided
they are registered to match the astrometric reference frame of the reference image.

For a given source, 4 background levels and their errors are estimated in each image from which the
user can choose: mode, mean, resistant mean and low mean.  The mode and resistant mean are calculated using
the MMM and RESISTANT\_MEAN routines, respectively, which can be found in the IDL Astronomy User's
Library \footnote{http://idlastro.gsfc.nasa.gov/contents.html}.  The low mean is found by using MEANCLIP
(also found in the IDL Astronomy User's Library) to compute a $3\sigma$ clipped mean of the
25\% lowest valued pixels in the background annulus.  In general, the mode is our preferred estimate
of the background level, however, the other estimates can provide a consistency
check or an alternative value if appropriate.  The estimated background value (per pixel) is
multiplied by the number of pixels in the aperture and then subtracted from the sum
of the pixels in the aperture to obtain the total background subtracted source counts.
In the case of the VLA images, the image units of Jy beam$^{-1}$ are first converted
to Jy pix$^{-1}$, where the conversion factor is given by the pixel area of
the 2-D Gaussian beam.  The background subtracted flux density (in Jy) is then computed
as described above.  

Error estimates include contributions from three sources: Poisson noise from the source
counts, Poisson noise from the background counts, and the error in measuring the
background level.  These error terms are calculated in units of electrons, added in quadrature
and converted back to counts using the gain.  A $1\sigma$ detection limit is found in the same
way, but only includes contributions from the Poisson noise from the background counts
and the error in measuring the background level.

\end{document}